\documentclass[sigconf,screen]{acmart}
\AtBeginDocument{%
  }

\usepackage{amsmath,amsfonts}
\allowdisplaybreaks[4]
\usepackage[linesnumbered,ruled,vlined]{algorithm2e}
\usepackage{makecell}
\usepackage{wrapfig}
\usepackage{graphicx}
\allowdisplaybreaks[4]
\usepackage{multirow}
\usepackage{colortbl}
\usepackage{textcomp}
\usepackage{xcolor}
\usepackage{pdfpages}
\usepackage{booktabs}
\usepackage{hyperref}
\usepackage{pifont}
\usepackage{subcaption}
\usepackage{enumitem}
\usepackage{orcidlink}
\usepackage{bbm}
\definecolor{tablecolor}{HTML}{C4D5B2}
\definecolor{cifarcolor}{HTML}{F3E6CA}
\definecolor{tinycolor}{HTML}{BCC6DD}
\definecolor{domaincolor}{HTML}{EFD6D1}

\copyrightyear{2026}
\acmYear{2026}
\setcopyright{cc}
\setcctype{by}
\acmConference[CIKM '26] {Proceedings of the 35th ACM International Conference on Information and Knowledge Management}{November 7--11, 2026}{Rome, Italy.}
\acmBooktitle{Proceedings of the 35th ACM International Conference on Information and Knowledge Management (CIKM '26), November 7--11, 2026, Rome, Italy}
\acmISBN{979-8-4007-2539-5/2026/11}
\acmDOI{10.1145/3799682.3840747}

\begin{document}

\title{When Is Shallow Enough? Adaptive Split Federated Learning with Client-Specific Sufficiency Estimation}



\author{Wenhao Yuan}
\authornote{Both authors contributed equally to this research.}
\orcid{0009-0001-6625-7496}
\affiliation{%
  \institution{The University of Hong Kong}
  \city{Hong Kong, SAR}
  \country{China}}
\email{wenhao.yuan@connect.hku.hk}

\author{Chenchen Lin}
\authornotemark[1]
\orcid{0009-0002-8473-6068}
\affiliation{%
  \institution{Sun Yat-sen University}
  \city{Zhuhai}
  \country{China}}
\email{linchch7@mail2.sysu.edu.cn}

\author{Wentao Hu}
\orcid{0000-0002-2071-9341}
\affiliation{%
  \institution{The Hong Kong Polytechnic University}
  \city{Hong Kong, SAR}
  \country{China}}
\email{wayne-wt.hu@connect.polyu.hk}

\author{Jian Chen}
\orcid{0000-0002-4570-2271}
\affiliation{%
  \institution{The University of Hong Kong}
  \city{Hong Kong, SAR}
  \country{China}}
\email{ccccccj03@connect.hku.hk}

\author{Jinfeng Xu}
\orcid{0009-0001-7876-3740}
\affiliation{%
  \institution{The University of Hong Kong}
  \city{Hong Kong, SAR}
  \country{China}}
\email{jinfeng@connect.hku.hk}

\author{Shujie Li}
\orcid{0000-0001-5239-2454}
\affiliation{%
  \institution{The University of Hong Kong}
  \city{Hong Kong, SAR}
  \country{China}}
\email{u3012850@connect.hku.hk}

\author{Edith Cheuk Han Ngai}
\orcid{0000-0002-3454-8731}
\authornote{Corresponding Author.}
\affiliation{%
  \institution{The University of Hong Kong}
  \city{Hong Kong, SAR}
  \country{China}}
\email{chngai@eee.hku.hk}

\renewcommand{\shortauthors}{Wenhao Yuan et al.}

\begin{abstract}
\textit{Split Federated Learning} (SFL) enables distributed model training by splitting networks between the server and clients. However, under client heterogeneity, the conventional static split strategy may be suboptimal because clients can differ in data distributions, adaptation dynamics, and representation learning progress, making a single split point insufficient to accommodate client-specific training states. In this paper, we propose \textsc{FedSGA}, a \textbf{S}ufficiency-\textbf{G}uided \textbf{A}daptive split \textbf{Fed}erated learning framework that addresses this question through client-specific shallow sufficiency estimation. First, we introduce a client-specific adaptation channel based on private prompt tokens, which tracks local adaptation dynamics separately from the shared backbone and provides a lightweight signal for detecting whether client adaptation remains active. To further avoid repeated online probing over multiple candidate depths, we design a shallow sufficiency estimator that combines cross-client semantic alignment, temporal interface stability, and prompt-state variation to estimate whether the shallowest split is already sufficient. Finally, we introduce a split-compatible interface harmonization module that projects activations from different split depths into a shared semantic space, improving the comparability of heterogeneous client interfaces before server-side prediction. Extensive experiments on multiple heterogeneous benchmarks demonstrate the effectiveness of \textsc{FedSGA} in improving model performance compared with state-of-the-art methods while reducing unnecessary client-side computation. 
\end{abstract}






\maketitle

\section{Introduction}

\textit{Federated Learning} (FL)~\cite{mcmahan2017communication} has emerged as a standard paradigm for collaborative model training across distributed clients without directly exposing their raw data. To further reduce client-side computation and support scalable deployment, \textit{Split Federated Learning} (SFL)~\cite{thapa2022splitfed} partitions the model into a client-side front submodel and a server-side back submodel, where clients transmit intermediate activations to the server for collaborative optimization. Existing SFL studies have mainly focused on resource-aware system designs, such as reducing computation, communication, or synchronization overhead~\cite{lin2025adaptsfl, ilhan2023scalefl, shen2023ringsfl, liao2024parallelsfl}. However, the role of data-induced heterogeneity in determining an appropriate split point remains less explored~\cite{dachille2025impact}. In particular, vanilla SFL commonly adopts a shared static split point for all clients, implicitly assuming that shallow representations evolve with comparable semantic maturity across heterogeneous clients. This assumption can break down under client heterogeneity, where different data distributions and adaptation dynamics lead to client-specific representation learning trajectories, making a fixed split point suboptimal for both optimization efficiency and model generalization.

In practical FL systems, client heterogeneity is not limited to resource diversity, but also arises from differences in data distributions and local learning behaviors~\cite{wang2025federated, chen2024fair, fu2025beyond}. Among these factors, \textit{domain skew} is a representative and practically important case: clients may collect data from different locations, sensors, or environments, resulting in discrepancies in feature distributions and semantic structures across clients~\cite{ye2023heterogeneous}. Such heterogeneity induces non-IID training behaviors and can substantially impair collaborative optimization and model generalization~\cite{huang2023rethinking}. In SFL, the impact of client heterogeneity is further amplified at the client-server interface. Since the server-side model is trained on intermediate activations produced by client-side front submodels, heterogeneous clients may expose interfaces with different semantic maturity, domain-dependent feature biases, or incompatible representation geometries. As a result, the shared server-side model has to integrate poorly aligned intermediate representations, which can destabilize collaborative training and degrade performance~\cite{thapa2022splitfed}.

While client-specific split adaptation appears necessary under heterogeneous SFL, existing adaptive split strategies are still limited in how they determine when and where to split. Resource-aware methods typically adjust split points according to device capability, latency, or communication cost, but such system-level criteria do not indicate whether the transmitted interface contains sufficient task-relevant information for server-side learning. Alternatively, online probing over multiple candidate depths can provide more direct evidence of split suitability, yet repeated depth evaluation introduces additional forward computation and becomes costly throughout training. More importantly, criteria based solely on the current interface representation can be unreliable under client heterogeneity: a shallow interface may appear semantically aligned at the current round, while the underlying client-local adaptation process remains far from stabilized.

These limitations expose two unresolved challenges in heterogeneous SFL: \textit{\textbf{\uppercase\expandafter{\romannumeral1}. Unreliable split decisions under client-specific adaptation:}} For each client, it remains unclear whether an early interface already preserves sufficient task-relevant information for server-side learning. Such reliability depends not only on the current representation quality, but also on whether the client-side model has reached a stable local adaptation state. Therefore, split decisions should account for both representation quality and client-specific adaptation dynamics without relying on exhaustive depth evaluation. \textit{\textbf{\uppercase\expandafter{\romannumeral2}. Heterogeneous interface compatibility:}} When clients offload from different depths, the server receives activations with different semantic granularity. Under client heterogeneity, these activations can be further affected by client-specific distribution shifts and adaptation states, making them difficult to integrate within a shared server-side model. These challenges raise the central question of this work: \textit{How can SFL adapt split points according to client-specific interface reliability while preserving compatible server-side learning across heterogeneous clients?}

To bridge this gap, we present a novel solution, a \textbf{S}ufficiency-\textbf{G}uided \textbf{A}daptive split \textbf{Fed}erated learning framework for heterogeneous SFL (\textsc{FedSGA}). For challenge~\textbf{\uppercase\expandafter{\romannumeral1}}, we first introduce a client-specific adaptation channel, as detailed in \S~\ref{prompt}, where private prompt tokens serve as lightweight local states to track whether client-side adaptation remains active. Building on this adaptation signal, we further develop a shallow sufficiency estimator in \S~\ref{adaptive_split}, which determines whether the shallowest interface is already sufficient for collaborative training by jointly considering cross-client semantic alignment, temporal interface stability, and prompt-state variation, without exhaustively probing all candidate split depths. For challenge~\textbf{\uppercase\expandafter{\romannumeral2}}, we introduce a split-compatible interface harmonization module in \S~\ref{alignment_collaboration}, which maps activations from heterogeneous split depths into a shared semantic space before server-side prediction. By conditioning the projection on split-depth information and regularizing the projected space with server-maintained class prototypes, \textsc{FedSGA} improves the comparability of heterogeneous client interfaces and stabilizes collaborative server-side learning. Our main contributions are summarized as follows:
\begin{itemize}
\item We identify the limitation of static split points in SFL under client heterogeneity and formulate adaptive split selection as a client-specific shallow sufficiency estimation problem. This perspective shifts split selection from resource-driven partitioning to sufficiency-guided decision making, where each client determines whether its shallow interface is reliable for collaborative server-side training.

\item We propose \textsc{FedSGA}, a sufficiency-guided adaptive split learning framework. \textsc{FedSGA} introduces a client-specific adaptation channel based on private prompt tokens to track local adaptation dynamics, and develops a shallow sufficiency estimator without exhaustive online probing over all candidate depths. We further design a split-compatible interface harmonization module that maps heterogeneous-depth activations into a shared semantic space with prototype-based alignment.

\item  We conduct extensive experiments under both general statistical heterogeneity and domain heterogeneity, including CIFAR-10, CIFAR-100, Tiny-ImageNet, and DomainNet, demonstrating the effectiveness of \textsc{FedSGA}.
\end{itemize}

\section{Related Works}

\subsection{Split Federated Learning}

Split Federated Learning (SFL) has been widely studied as a collaborative training paradigm that partitions a model between clients and the server, where intermediate activations and gradients are exchanged across the split interface~\cite{han2024convergence, liao2024parallelsfl, lin2025hierarchical, gao2024pipesfl, djuhera2025r}. By moving part of the computation to the server, SFL reduces the local training burden on clients while preserving the distributed nature of federated optimization. Existing studies mainly improve SFL from a system perspective, including privacy evaluation~\cite{wu2024evaluating}, communication reduction~\cite{shiranthika2025splitfedzip}, pipeline execution~\cite{gao2024pipesfl, liao2024parallelsfl}, hierarchical training~\cite{lin2025hierarchical}, and acceleration under resource-constrained deployment~\cite{lin2025adaptsfl, xu2024accelerating, tian2024breaking}. Other efforts further consider client-server workload balancing or heterogeneous device capabilities to reduce training overhead and improve practical scalability~\cite{yao2025pairingfl, fan2025madrl}. Beyond system efficiency, personalized SFL introduces client-specific components to improve local adaptation under statistical heterogeneity~\cite{dai2025psfl, pervej2025personalized, zheng2024ppsfl, xie2025tackling}. Some studies further regularize intermediate representations or improve client-server collaboration through feature alignment and distillation objectives~\cite{liao2024mergesfl, luo2024federated, mao2024towards}. These efforts indicate that the split interface is not merely a communication boundary, but also affects how client-side representations are exposed to the shared server-side model. Nevertheless, most existing approaches still rely on fixed split structures or resource-oriented partition strategies, leaving the representation-level role of the split interface less explicitly studied.

\begin{figure*}[t]
\centerline{\includegraphics[width=1.0\textwidth, trim=0 0 0 0,clip]{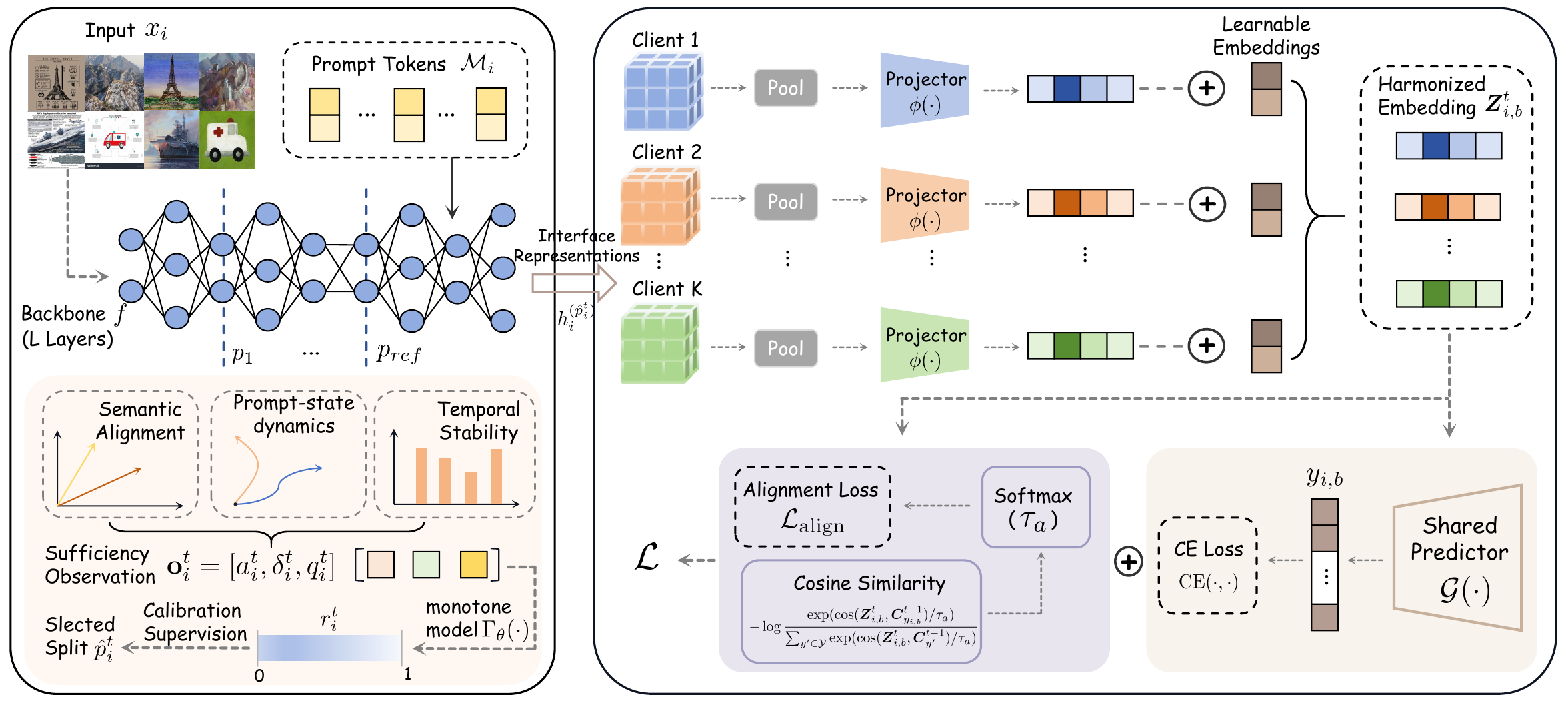}}
\caption{\textbf{Overview illustration} of our proposed \textsc{FedSGA} method.}
\label{framework}
\Description{framework}
\end{figure*}

\subsection{Adaptive Split Learning with Heterogeneity}
Client heterogeneity has been extensively studied in FL, where clients may differ in data distributions, domain semantics, and local optimization behaviors~\cite{gai2025mufti, gao2025multimodal, wang2024taming, orouji2024domain}. Such heterogeneity can cause client drift, unstable optimization, and degraded generalization~\cite{jia2024dapperfl}. Existing approaches mainly improve robustness under heterogeneous data through domain-invariant representation learning~\cite{li2024fedcir}, feature distribution alignment~\cite{gupta2025fedalign}, update regularization~\cite{li2025federated, wang2025federated}, domain generalization~\cite{zhang2023federated, bai2024diprompt}, and domain adaptation~\cite{shenaj2023learning, liu2024ufda, feng2024robust}. Other studies exploit multi-source training~\cite{wang2024multi, wei2024multi}, adversarial distribution alignment~\cite{gupta2025fedalign, zhou2024federated}, or domain disentanglement~\cite{bai2024diprompt, chen2024disentanglement} to reduce cross-domain discrepancies. While effective in standard FL, these methods typically assume a unified model architecture and do not explicitly consider how client heterogeneity affects the semantic quality of split interfaces. \textit{Adaptive split learning} provides a natural way to relax the rigidity of static split architectures by allowing different clients to offload at different depths~\cite{lin2025adaptsfl, ilhan2023scalefl, shen2023ringsfl, liao2024parallelsfl}. Existing adaptive split strategies, however, are often driven by resource-side factors such as computation budget, latency, communication overhead, or device capability~\cite{lin2025adaptsfl, xu2024accelerating, fan2025madrl}. These criteria are useful for system efficiency but do not directly indicate whether a shallow interface is semantically sufficient for server-side learning. More direct depth evaluation through online probing can better reflect split suitability, but repeated probing over multiple candidate depths introduces non-negligible overhead~\cite{shan2026splitft, zhang2025ampere}. Moreover, under client heterogeneity, representation-level signals alone may be unreliable because an interface that appears aligned at the current round can still be produced by an actively adapting client.

\subsection{Representation Alignment in FL}
Representation alignment has been widely studied in FL to improve cross-client consistency under heterogeneous data distributions. Existing methods reduce representation discrepancy through feature alignment, prototype regularization, contrastive learning, or knowledge distillation~\cite{gupta2025fedalign, liao2024mergesfl, mao2024towards, luo2024federated}. Prototype-based approaches leverage class-level semantic anchors to encourage consistent feature geometry across clients~\cite{li2024fedcir, zhou2024federated}, while contrastive and distillation-based methods improve feature transferability by aligning latent representations or prediction distributions~\cite{mao2024towards, luo2024federated}. Other studies further incorporate domain-aware alignment objectives to mitigate feature shifts induced by heterogeneous environments~\cite{gupta2025fedalign, bai2024diprompt, chen2024disentanglement}. Although these methods improve representation consistency in conventional FL, most of them assume homogeneous network structures or feature spaces across clients. In adaptive SFL, however, clients may offload activations from different backbone depths, resulting in heterogeneous semantic granularity at the split interface~\cite{lin2025adaptsfl, liao2024parallelsfl}. Under such heterogeneous split configurations, directly aligning representations across clients becomes substantially more challenging, since the transmitted activations may differ not only in domain semantics but also in abstraction level~\cite{dachille2025impact}. Existing federated learning alignment methods are therefore insufficient for adaptive split learning, where representation alignment must additionally account for heterogeneous semantic granularity induced by client-specific split depths.

\section{Methodology}
In this section, we present \textsc{FedSGA}, a framework for adaptive split learning under client heterogeneity, which formulates split selection as a client-specific shallow sufficiency estimation problem. The framework is motivated by the insight that the quality of a shallow interface depends not only on its current representation quality, but also on whether client-local adaptation has sufficiently stabilized. We begin with the preliminaries and notation in \S~\ref{problem_formulation} and introduce a client-specific adaptation channel in \S~\ref{prompt}, where private prompt tokens track local adaptation dynamics separately from the shared backbone. We then present a shallow sufficiency estimator in \S~\ref{adaptive_split}, which combines semantic alignment, temporal stability, and prompt-state variation to estimate whether the shallowest split is already sufficient for collaborative training. Finally, \S~\ref{alignment_collaboration} introduces a split-compatible interface harmonization module that projects activations from different split depths into a shared semantic space before server-side prediction.

\subsection{Preliminaries and Notation} \label{problem_formulation}

\textit{\textbf{Split Federated Learning.}} The global model is partitioned across the client and server sides, with training performed collaboratively through intermediate activations and gradients. We denote the model parameters by $\boldsymbol{w} = (\boldsymbol{w}_{s}, \{\boldsymbol{w}_{i}\}_{i \in \mathcal{I}})$, where $\boldsymbol{w}_{s}$ is the server-side submodel shared across clients, and $\boldsymbol{w}_{i}$ is the client-side front submodel maintained by client $i \in \mathcal{I} = \{1, \ldots, N\}$. Each client $i$ owns a private dataset $\mathcal{D}_{i}$ with datasize $D_{i}$. At global iteration $t \in \{0, \ldots, T\}$, client $i$ computes the forward pass up to the split point and sends the resulting interface activations to the server. The server completes the remaining forward and backward computation on $\boldsymbol{w}_{s}$, and \emph{returns the interface gradients} to the client to update $\boldsymbol{w}_{i}$. The server then updates the shared submodel $\boldsymbol{w}_{s}$ using the contributions from participating clients and broadcasts the updated parameters for the next iteration. Let $\mathcal{B}_{i}$ denote the batch sampled from $\mathcal{D}_{i}$. The training objective is formulated as
\begin{align}
\min_{\boldsymbol{w}_{s}, \{\boldsymbol{w}_{i}\}} F(\boldsymbol{w}) = \sum\nolimits_{i \in \mathcal{I}} \frac{D_{i}}{\sum\nolimits_{j \in \mathcal{I}} D_{j}} \mathbb{E}_{\mathcal{B}_{i} \sim \mathcal{D}_{i}} [F_{i} (\boldsymbol{w}_{s}, \boldsymbol{w}_{i}; \mathcal{B}_{i})].
\end{align}

\textit{\textbf{Client Heterogeneity and Domain Skew.}} In heterogeneous federated learning scenarios, clients generally exhibit substantial heterogeneity in data distributions, optimization behavior, and representation learning progress, often alongside diverse computational capacities. Such heterogeneity directly affects the semantic quality of client-side interfaces and the appropriateness of a fixed split point for collaborative training. Domain skew is a common and practically important instance of client heterogeneity, where clients share the same label space but differ in their class-conditional input distributions. Formally, with $\mathbb{P}_{m}$ and $\mathbb{P}_{n}$ denoting the data distributions on clients $m$ and $n$, domain skew is characterized by $\mathbb{P}_{m}(x|y) \ne \mathbb{P}_{n}(x|y)$. Under this configuration, client-specific feature characteristics may lead to different representation learning trajectories across clients, making the suitability of a split point inherently client-dependent.

\subsection{Client-Specific Adaptation Channel} \label{prompt}
Under client heterogeneity, interface representations in split learning alone may not fully reflect the local adaptation status of each individual client. For example, even when clients share the same label space, their class-conditional feature distributions may vary, and local adaptation processes may progress at different rates. Therefore, a shallow representation may already appear semantically aligned or temporally stable even when the client is still undergoing non-negligible local adaptation.

To address this issue, we introduce a client-specific adaptation channel decoupled from the shared backbone. Instead of tracking the full feature space or globally shared parameters, we maintain a small set of client-specific prompt tokens for each client. Note that these prompt parameters are updated only via the local objective of the corresponding client and are not directly involved in cross-client aggregation. In contrast to shared backbone parameters, whose updates are entangled with global optimization, or interface features, which can be noisy and strongly batch-dependent, the evolution of prompt tokens provides a cleaner view of client-local adaptation. The prompts are not introduced for test-time personalization. Rather, they serve as a lightweight private state for monitoring whether client-specific adaptation remains active.

Let $f(\cdot)$ denote a backbone with $L$ layers. For each client $i \in \mathcal{I}=\{1,\ldots,N\}$, we maintain a set of client-specific prompt tokens $\mathcal{M}_i = \{\boldsymbol{\epsilon}_{i,1}, \ldots, \boldsymbol{\epsilon}_{i,m}\}$, $\boldsymbol{\epsilon}_{i,k} \in \mathbb{R}^{d}$, where $m$ is the number of prompt tokens and $d$ is the embedding dimension. Given an input sequence $x=[x_{\mathrm{cls}},x_1,\ldots,x_n]$, we prepend the prompts to obtain the client-conditioned input $\tilde{x}_i = [x_{\mathrm{cls}}, \boldsymbol{\epsilon}_{i,1}, \ldots, \boldsymbol{\epsilon}_{i,m}, x_1,\ldots,x_n]$. The backbone then processes the resulting sequence. Let $f^{(1:l)}(\cdot)$ denote the transformation up to layer $l$, and define $h_i^{(1:l)} = f^{(1:l)}(\tilde{x}_i)$ as the corresponding intermediate representation. To obtain a compact summary of the prompt state at communication round $t$, let $\ell_i^t$ denote the local training loss of client $i$. For each prompt token $\boldsymbol{\epsilon}_{i,k}$, we compute a first-order contribution score $c_{i,k}^t = |(\partial \ell_i^t / \partial \boldsymbol{\epsilon}_{i,k})^{\top} \boldsymbol{\epsilon}_{i,k}|$, which measures the sensitivity of the local objective along the corresponding prompt direction. A larger value indicates that the prompt token is more actively involved in the current local update. \footnote{Note that the gradients are available during standard backpropagation, thus calculating the contribution scores introduces negligible additional overhead.} Then, we aggregate the prompt tokens into an importance-weighted adaptation descriptor as follows:
\begin{align}
\boldsymbol{m}_i^t = \sum\nolimits_{k=1}^{m} \alpha_{i,k}^t \boldsymbol{\epsilon}_{i,k}, \  \alpha_{i,k}^t = \frac{\exp(c_{i,k}^t/\tau_m)}{\sum\nolimits_{j=1}^{m}\exp(c_{i,j}^t/\tau_m)},
\label{prompt_descriptor}
\end{align}
where $\tau_m$ is a temperature parameter. This weighting emphasizes prompt directions that are more influential to the current local objective and are therefore more indicative of ongoing client-specific adaptation. The vector $\boldsymbol{m}_i^t$ serves as a compact descriptor of the client-specific adaptation channel. Since it evolves jointly with the prompt parameters, its temporal variation provides a low-cost signal that reflects changing client-specific adaptation patterns during local training.

\subsection{Shallow Sufficiency Estimation} \label{adaptive_split}

Building on the client-specific adaptation descriptor in \S~\ref{prompt}, we formulate adaptive split selection as a \emph{client-specific shallow sufficiency estimation} problem. Existing adaptive split strategies typically rely either on explicit online probing over multiple candidate depths or on hand-crafted depth-specific criteria. Under client heterogeneity, both choices are unsatisfactory. Exhaustive probing incurs non-trivial runtime overhead, while fixed criteria can be brittle when clients differ in data distributions, local adaptation dynamics, or representation learning progress. Instead of repeatedly evaluating all candidate depths online, each client estimates whether the shallowest candidate interface already contains sufficient task-relevant information for collaborative server-side training.

Let $\mathcal{P}=\{p_1,p_2,\ldots,p_K\}$ denote the set of candidate split depths aligned with backbone block boundaries, where $p_1$ is the shallowest candidate and $p_K$ is the deepest one. For client $i$ at communication round $t$, the representation produced at depth $p$ is $h_i^{(p,t)}(\tilde{x}_i) = f_t^{(1:p)}(\tilde{x}_i)$, where $\tilde{x}_i$ is the prompt-conditioned input. We use $p_1$ as the reference shallow interface. If it already preserves sufficient task-relevant information for the server-side model, additional local forwarding is unnecessary; otherwise, the client should retain a deeper local prefix before offloading. We operationalize shallow sufficiency by comparing the shallow reference split with a fixed deeper reference split. Specifically, we select a single moderately deep reference depth $p_{\mathrm{ref}} \in \mathcal{P}$ and use it consistently across clients and training rounds. At selected calibration rounds, client $i$ evaluates the probe losses at $p_1$ and $p_{\mathrm{ref}}$, denoted by $\mathcal{L}_i^{(p_1,t)}$ and $\mathcal{L}_i^{(p_{\mathrm{ref}},t)}$, respectively. The shallow interface is regarded as sufficient if $\mathcal{L}_i^{(p_1,t)} \leq (1+\beta) \mathcal{L}_i^{(p_{\mathrm{ref}},t)}$, where $\beta$ is a relative tolerance margin. This criterion is not intended to identify the globally optimal split at every round. Instead, it provides a stable and low-cost operational target for learning whether the shallowest candidate is already acceptable relative to a fixed deeper reference.

To avoid repeated online depth comparison, we predict this sufficiency event from lightweight observable signals. We first measure whether the shallow interface is aligned with cross-client class semantics. Let $\bar{h}_i^{(p_1,t)}(\tilde{x}_i)$ denote the pooled shallow representation, and let $\phi_s(\cdot)$ be a lightweight projection head used only for sufficiency estimation. Given the server-maintained class prototypes $\{\boldsymbol{C}_{y}^{t-1}\}_{y\in\mathcal{Y}}$, we define
\begin{align}
a_i^t = \frac{1}{|\mathcal{B}_i^t|} \sum\nolimits_{x\in\mathcal{B}_i^t} \cos(\phi_s(\bar{h}_i^{(p_1,t)}(\tilde{x}_i)), \boldsymbol{C}_{y(x)}^{t-1}),
\end{align}
where a larger $a_i^t$ indicates that the shallow representation is more consistent with the global class-level geometry and is therefore more likely to support accurate server-side prediction. However, semantic alignment alone is insufficient under heterogeneous client adaptation. A shallow representation may appear well aligned with the current prototypes even when the client is still undergoing substantial local adaptation. Therefore, we incorporate the prompt-state dynamics introduced in \S~\ref{prompt}. Specifically, we define
\begin{align}
q_i^t = \|\boldsymbol{m}_i^t - \boldsymbol{m}_i^{t-1}\|_2,
\end{align}
where $\boldsymbol{m}_i^t$ is the importance-weighted prompt descriptor at $t$. Since the prompt tokens are updated only through client-local supervision, a large $q_i^t$ indicates that the client-specific adaptation state is still changing substantially. Thus, $q_i^t$ serves as a compact observation of residual client-local adaptation. It is not intended to estimate the client domain directly; rather, it complements representation-level statistics by indicating whether the local adaptation process remains unresolved. We further measure the temporal stability of the shallow interface by
\begin{align}
\delta_i^t = \|H_i^t - H_i^{t-1}\|_2, \  H_i^t = \frac{1}{|\mathcal{B}_i^t|}\sum\nolimits_{x\in\mathcal{B}_i^t}\bar{h}_i^{(p_1,t)}(\tilde{x}_i),
\end{align}
where a smaller $\delta_i^t$ suggests that the shallow representation has entered a more stable regime across communication rounds. We then construct the sufficiency observation as $\mathbf{o}_i^t = [a_i^t,\delta_i^t,q_i^t]$. Based on this observation, the client estimates the probability that the shallow interface is sufficient:
\begin{align}
r_i^t = \Gamma_\theta(\mathbf{o}_i^t,r_i^{t-1}),
\end{align}
where $r_i^t\in[0,1]$ denotes the estimated shallow-sufficiency probability. We implement $\Gamma_\theta(\cdot)$ as a low-capacity monotone model whose output increases with prototype alignment and decreases with representation drift and prompt-state variation, imposing a conservative inductive bias: a shallow interface should be considered reliable only when it is semantically aligned, temporally stable, and not accompanied by strong residual client-local adaptation.

\begin{algorithm}[t]
\caption{Shallow Sufficiency Calibration}
\label{fedSGA_calibration}
\KwIn{$\mathcal{B}_i^t$; observation $\mathbf{o}_i^t$; previous estimate $r_i^{t-1}$; splits $p_1,p_{\mathrm{ref}}$; tolerance $\beta$; estimator $\Gamma_\theta$.}
\KwOut{Updated estimator $\Gamma_\theta$.}

Evaluate probe losses $\mathcal{L}_i^{(p_1,t)}$ and $\mathcal{L}_i^{(p_{\mathrm{ref}},t)}$ on $\mathcal{B}_i^t$\;

\If{$\mathcal{L}_i^{(p_1,t)} \leq (1+\beta)\mathcal{L}_i^{(p_{\mathrm{ref}},t)}$}{
  $s_i^t\leftarrow 1$\;
}
\Else{
  $s_i^t\leftarrow 0$\;
}

Compute $r_i^t \leftarrow \Gamma_\theta(\mathbf{o}_i^t,r_i^{t-1})$\;
Store the calibration pair $((\mathbf{o}_i^t,r_i^{t-1}),s_i^t)$\;
Update $\Gamma_\theta$ using accumulated calibration pairs\;

\Return $\Gamma_\theta$\;
\end{algorithm}

The estimator is then trained using low-frequency calibration supervision. Let $\mathcal{T}_{\mathrm{cal}}\subseteq\{1,\ldots,T\}$ denote the set of calibration rounds. For each $t\in\mathcal{T}_{\mathrm{cal}}$, the client compares $p_1$ with the fixed deeper reference split $p_{\mathrm{ref}}$ and constructs the binary target
\begin{align}
s_i^t = \mathbbm{1} [\mathcal{L}_i^{(p_1,t)} \leq (1+\beta) \mathcal{L}_i^{(p_{\mathrm{ref}},t)}],
\end{align}
where $\Gamma_\theta$ is optimized to predict $s_i^t$ from $(\mathbf{o}_i^t,r_i^{t-1})$. This calibration is used only to provide sparse supervision during training and is not part of the online split decision at every round. Once the estimator is learned, the client can avoid explicit depth comparison and rely on the estimated sufficiency probability for split selection. Finally, with $\rho$ controlling the system-level trade-off between early offloading and local computation, we convert the sufficiency estimate into a discrete split depth through a budget-aware monotone policy as $\hat{p}_i^t = \Psi(r_i^t;\rho)$. Then, we instantiate $\Psi(\cdot)$ as
\begin{align}
\hat{p}_i^t = p_k, \  k = \min \{K, \max \{1, \lceil K(1-r_i^t+\rho) \rceil\}\},
\end{align}
where a larger $\rho$ shifts the policy toward deeper local computation, whereas a smaller $\rho$ favors earlier offloading. The client then computes the backbone only up to $\hat{p}_i^t$ and transmits $\boldsymbol{z}_i^t(\tilde{x}_i) = h_i^{(\hat{p}_i^t,t)}(\tilde{x}_i)$ to the server. This formulation replaces repeated online depth comparison with calibration-supervised sufficiency estimation that jointly considers cross-client semantic alignment, shallow-interface stability, and prompt-based client-local adaptation dynamics, enabling low-cost split selection while reducing premature shallow splitting across clients. The calibration procedure for learning the shallow-sufficiency estimator is summarized in Alg.~\ref{fedSGA_calibration}.

\begin{algorithm}[t]
\caption{Overall Training Procedure of \textsc{FedSGA}}
\label{fedSGA_overall}
\KwIn{$\mathcal{I}$; $T$; candidate splits $\mathcal{P}$; reference split $p_{\mathrm{ref}}$; calibration schedule $\mathcal{T}_{\mathrm{cal}}$; $\mu, \rho,\lambda,\tau_m,\tau_a,\beta$.}
\KwOut{Trained $\boldsymbol{w}_s$ and $\{\boldsymbol{w}_i\}_{i\in\mathcal{I}}$.}

\textbf{Initialize} $\boldsymbol{w}_s^0$, $\{\boldsymbol{w}_i^0\}$, $\{\mathcal{M}_i^0\}$, $\Gamma_\theta$, calibration buffer, and class prototypes $\{\boldsymbol{C}_y^0\}_{y\in\mathcal{Y}}$\;

\For{$t=1$ \KwTo $T$}{
  Sample participating clients $\mathcal{S}_t\subseteq\mathcal{I}$ and broadcast $\boldsymbol{w}_s^{t-1}$, $\{\boldsymbol{C}_y^{t-1}\}_{y\in\mathcal{Y}}$\;

  \ForEach{$i\in\mathcal{S}_t$}{
    Sample $\mathcal{B}_i^t\sim\mathcal{D}_i$ and construct prompt-conditioned inputs $\tilde{x}_i$\;
    Compute $\boldsymbol{m}_i^t$ by \eqref{prompt_descriptor} and form $\mathbf{o}_i^t=[a_i^t,\delta_i^t,q_i^t]$\;

    \If{$t\in\mathcal{T}_{\mathrm{cal}}$}{
      Update $\Gamma_\theta$ using Alg.~\ref{fedSGA_calibration} with $\mathcal{B}_i^t$, $\mathbf{o}_i^t$, and $r_i^{t-1}$\;
    }

    Estimate $r_i^t=\Gamma_\theta(\mathbf{o}_i^t,r_i^{t-1})$ and select $\hat{p}_i^t=\Psi(r_i^t;\rho)$\;
    Compute $\boldsymbol{z}_i^t=h_i^{(\hat{p}_i^t,t)}(\tilde{x}_i)$ and upload $(\boldsymbol{z}_i^t,\hat{p}_i^t,\{y_b\})$ to the server\;
  }

  Project uploaded interfaces into $\{\boldsymbol{Z}_{i,b}^t\}$ using the depth-aware harmonization module\;
  Compute $\mathcal{L}_{\mathrm{align}}$ and $\mathcal{L}$ by \eqref{align_loss}--\eqref{overall_loss}\;
  Update $\boldsymbol{w}_s$, $\phi$, and $\mathcal{G}$, and return interface gradients to clients\;
  Clients update $\{\boldsymbol{w}_i,\mathcal{M}_i\}_{i\in\mathcal{S}_t}$; server updates $\{\boldsymbol{C}_y^t\}_{y\in\mathcal{Y}}$\;
}
\Return $\boldsymbol{w}_s^T$ and $\{\boldsymbol{w}_i^T\}_{i\in\mathcal{I}}$\;
\end{algorithm}

\subsection{Split-Compatible Interface Harmonization}  \label{alignment_collaboration}

The proposed adaptive split mechanism allows clients to offload representations at different backbone depths. As a result, the server receives interface representations with heterogeneous semantic granularity. Under client heterogeneity, this depth heterogeneity can be further coupled with client-specific distribution shift or different adaptation states, making the received interfaces less directly comparable across clients. To stabilize joint server-side training, we introduce a lightweight harmonization module that maps interface representations from different split depths into a shared semantic space before prediction.

For an input sample $\tilde{x}_{i,b} \in \mathcal{B}_i^t$, client $i$ transmits the interface representation $\boldsymbol{z}_{i,b}^t = h_i^{(\hat{p}_i^t,t)}(\tilde{x}_{i,b})$, where $\hat{p}_i^t$ is the split depth selected by the shallow sufficiency estimator. We first obtain a compact interface feature by applying a pooling operator $\tilde{\boldsymbol{z}}_{i,b}^t = \mathrm{Pool}(\boldsymbol{z}_{i,b}^t)$, where $\mathrm{Pool}(\cdot)$ denotes token- or spatial-level aggregation, such as average pooling. To improve compatibility across heterogeneous split depths and heterogeneous client representations, we apply a lightweight depth-aware projector $\boldsymbol{Z}_{i,b}^t = \phi(\tilde{\boldsymbol{z}}_{i,b}^t, \boldsymbol{e}(\hat{p}_i^t))$, where $\phi(\cdot)$ is a shared projection module and $\boldsymbol{e}(\hat{p}_i^t)$ is a learnable embedding of the selected split depth. The depth embedding provides explicit information about the semantic level at which the representation is produced. Conditioning the projector on this depth information allows it to compensate for systematic shifts induced by different split locations, making representations from different depths compatible for joint optimization.

We further regularize the projected space using server-maintained class prototypes. Specifically, we introduce the following prototype-based alignment objective:
\begin{align} \label{align_loss}
\mathcal{L}_{\mathrm{align}} = -\sum\nolimits_{i\in\mathcal{I}} \sum\nolimits_{b\in\mathcal{B}_i^t} \log\frac{\exp(\cos(\boldsymbol{Z}_{i,b}^t,\boldsymbol{C}_{y_{i,b}}^{t-1})/\tau_a)}{\sum\nolimits_{y'\in\mathcal{Y}} \exp(\cos(\boldsymbol{Z}_{i,b}^t,\boldsymbol{C}_{y'}^{t-1})/\tau_a)},
\end{align}
where $\tau_a$ is a temperature parameter and $\boldsymbol{C}_{y}^{t-1}$ denotes the prototype of class $y$ from the previous communication round. Since each prototype aggregates class-level information across participating clients, it provides a cross-client semantic anchor that remains useful under client heterogeneity. This regularizer encourages projected representations from different split depths and different client states to preserve a consistent class-level geometry. The overall server-side objective is then given by
\begin{align} \label{overall_loss}
\mathcal{L} = \sum\nolimits_{i\in\mathcal{I}} \sum\nolimits_{b\in\mathcal{B}_i^t} \mathrm{CE} (\mathcal{G}(\boldsymbol{Z}_{i,b}^t), y_{i,b}) + \lambda \mathcal{L}_{\mathrm{align}},
\end{align}
where $\mathcal{G}(\cdot)$ denotes the server-side prediction head and $\lambda$ balances the task loss and the alignment regularizer. After each server update, the class prototype is updated by an exponential moving average $\boldsymbol{C}_y^t \leftarrow \mu \boldsymbol{C}_y^{t-1} + (1-\mu) \frac{\sum\nolimits_{i\in\mathcal{S}_t}\sum\nolimits_{b\in\mathcal{B}_i^t}\mathbbm{1}[y_{i,b}=y]\boldsymbol{Z}_{i,b}^t}{\sum\nolimits_{i\in\mathcal{S}_t}\sum\nolimits_{b\in\mathcal{B}_i^t}\mathbbm{1}[y_{i,b}=y]}$, $\boldsymbol{C}_y^t \leftarrow \boldsymbol{C}_y^t/\|\boldsymbol{C}_y^t\|_2$. This module is not intended to fully eliminate the mismatch caused by heterogeneous split depths and client-specific adaptation states. Instead, it serves as a minimal harmonization layer that improves the stability of adaptive split training by increasing the comparability of interfaces produced by different split depths and heterogeneous client states. The overall training procedure of \textsc{FedSGA} is summarized in Alg.~\ref{fedSGA_overall}.

\subsection{Discussion}

\paragraph{Calibration Efficiency.}
\textsc{FedSGA} reduces the decision overhead of adaptive split selection by replacing per-round multi-depth probing with sparse two-depth calibration. A probing-based strategy that evaluates all $K$ candidate depths in $\mathcal{P}$ over $T$ rounds incurs $\mathcal{O}(KT)$ additional depth-comparison cost. In contrast, \textsc{FedSGA} uses only the comparison between $p_1$ and $p_{\mathrm{ref}}$ to calibrate the sufficiency estimator at selected rounds, and then selects $\hat{p}_i^t$ from $\mathcal{P}$ leveraging the estimated sufficiency probability $r_i^t$ during the ordinary training. This design does not eliminate calibration; rather, it amortizes the cost of depth comparison over training and avoids repeated exhaustive probing.

\paragraph{Client-side Computation.}
The selected split $\hat{p}_i^t$ specifies how much of the backbone is executed on client $i$. Since a larger sufficiency estimate $r_i^t$ induces a shallower split through the policy $\Psi(\cdot)$, clients whose shallow interfaces are estimated to be sufficient can offload earlier and avoid unnecessary deeper local computation. Therefore, the computational benefit of \textsc{FedSGA} comes from client-specific early offloading, rather than from uniformly enforcing the shallowest split for all clients. This distinction is important under client heterogeneity, where aggressive shallow splitting may reduce local computation but degrade the quality of transmitted interfaces.

\paragraph{Overall Complexity.}
The auxiliary computation introduced by \textsc{FedSGA} remains lightweight. After standard backpropagation provides prompt gradients, constructing the prompt descriptor only requires $\mathcal{O}(m)$ operations over $m$ prompt tokens. The sufficiency estimator $\Gamma_\theta$ is evaluated once per client in ordinary rounds, adding only a small prediction overhead. Sparse calibration compares only two depths, $p_1$ and $p_{\mathrm{ref}}$, over $|\mathcal{T}_{\mathrm{cal}}|$ calibration rounds, resulting in $\mathcal{O}(|\mathcal{T}_{\mathrm{cal}}|)$ probing cost rather than $\mathcal{O}(KT)$ exhaustive depth evaluation. On the server side, prototype alignment computes similarities between projected features and class prototypes, with $\mathcal{O}(B|\mathcal{Y}|)$ cost per client batch. Thus, the added cost of \textsc{FedSGA} is dominated by lightweight prompt summarization, sparse calibration, and compact prototype alignment, while the adaptive split policy can reduce the much larger client-side backbone computation.

\section{Numerical Experiments}\label{Numerical_Experiments}
In this section, we introduce the experimental setups and aim to answer the following research questions (RQs):
\begin{itemize}[itemsep=0pt, leftmargin=*, align=right]
\item \textbf{RQ1}: How does \textsc{FedSGA} perform compared with state-of-the-art FL and SFL methods under heterogeneous environments?
\item \textbf{RQ2}: How does each proposed component contribute to the overall performance of \textsc{FedSGA}?
\item \textbf{RQ3}: How sensitive to key hyperparameter configurations?
\item \textbf{RQ4}: Can \textsc{FedSGA} reduce client-side computational cost while maintaining strong collaborative learning performance?
\item \textbf{RQ5}: How effective is sufficiency-guided adaptive split selection compared with other split strategies?
\end{itemize}

\subsection{Experiment Setups}

\subsubsection{Datasets and Local Architecture.} 
We conduct our experiments on four typical and real-world datasets with different types and sorts, including CIFAR-10, CIFAR-100~\cite{krizhevsky2009learning}, Tiny-ImageNet~\cite{le2015tiny}, and DomainNet~\cite{peng2019moment}. These datasets cover both general statistical heterogeneity and domain-level heterogeneity, allowing us to examine the effectiveness of adaptive split learning under different client distribution shifts. For CIFAR-10, CIFAR-100, and Tiny-ImageNet, we construct IID partitions by uniformly assigning training samples to clients, and construct Non-IID partitions using the Dirichlet distribution~\cite{hsu2019measuring} with concentration parameter $0.1$. A smaller Dirichlet parameter induces stronger label distribution skew, which is used to simulate highly heterogeneous federated environments. For DomainNet, the same partitioning protocol is applied within its multi-domain data, so the evaluation reflects both domain-level visual variation and client-level Non-IID skew. We conduct experiments with two backbone architectures, ResNet-18~\cite{he2016deep} and ViT-B/16~\cite{dosovitskiy2020image}, to verify the generality of \textsc{FedSGA} across convolutional and transformer-based models. All compared methods use the same dataset partitions, backbone architectures, and training protocol for fair comparison.

\subsubsection{Baselines.}
We compare our method with the following two types of baselines: (\lowercase\expandafter{\romannumeral1})
\textit{Federated Learning Baselines.}
\textsc{FedAvg}~\citep{mcmahan2017communication} is the classical federated optimization framework based on iterative local training and global parameter averaging. \textsc{FedProx}~\citep{li2020federated} improves optimization stability under heterogeneous systems by introducing a proximal regularization term during local updates. \textsc{FedBABU}~\citep{oh2022fedbabu} decouples backbone and classifier optimization to improve personalization capability, while \textsc{FedGH}~\citep{yi2023fedgh} enhances federated representation learning through global prototype-guided classifier optimization. \textsc{FedAS}~\citep{yang2024fedas} addresses client inconsistency and straggler effects by jointly aligning local features and adaptively aggregating global parameters. (\lowercase\expandafter{\romannumeral2}) \textit{Split Federated Learning Baselines.} \textsc{SplitFed}~\cite{thapa2022splitfed} is a standard SFL framework that combines split learning with federated aggregation by partitioning the model between clients and the server and exchanging intermediate activations and gradients across the split interface. \textsc{MergeSFL}~\cite{liao2024mergesfl} reduces split learning overhead by merging intermediate representations during collaborative training. \textsc{FedMut}~\cite{hu2024fedmut} introduces mutual learning strategies to improve representation consistency between client and server models. \textsc{MU-SplitFed}~\cite{liang2025towards} improves communication efficiency in SFL through multi-user collaborative split learning. \textsc{MultiSFL}~\cite{xia2025multisfl} dynamically adjusts split strategies according to heterogeneous client resources and serves as the primary adaptive SFL baseline in our experiments.

\begin{table}[t]
\centering
\caption{Basic Information of Datasets}
\vspace{-8pt}
\renewcommand\arraystretch{0.8}
\resizebox{0.48\textwidth}{!}{
\begin{tabular}{c|c|c|c|c}
\toprule[1pt]
\textbf{Datasets}&\textbf{Training Size}&\textbf{Test Size}&\textbf{Class} &\textbf{Image Size}  \\ \cmidrule[0.5pt](l{1pt}r{0pt}){1-5}

CIFAR-10 & 50,000 & 10,000  & 10 & 3 $\times$ 32 $\times$ 32  \\ \cmidrule[0.5pt](l{1pt}r{0pt}){1-5}

CIFAR-100 & 50,000 & 10,000  & 100 & 3 $\times$ 32 $\times$ 32  \\ \cmidrule[0.5pt](l{1pt}r{0pt}){1-5}

Tiny-ImageNet & 100,000 & 10,000  & 200 & 3 $\times$ 64 $\times$ 64  \\ \cmidrule[0.5pt](l{1pt}r{0pt}){1-5}

DomainNet & 586,575 & 34,000  & 345 & 3 $\times$ 224 $\times$ 224  \\
\bottomrule[1pt]
\end{tabular}}
\label{basic_information_of_datasets}
\end{table}

\subsubsection{Hyperparameter settings}
We implement SGD as the default optimizer for local client-side training and report the average results over multiple runs. Unless otherwise specified, the number of clients is set to $N=20$, the local training epoch is set to $L=5$, and the batch size is set to 64. All clients participate in each communication round. For CIFAR-10, we use a learning rate of 0.1 and train the model for $T=100$ global rounds. For CIFAR-100, Tiny-ImageNet, and DomainNet, we use a learning rate of 0.01 and train for $T=300$ global rounds. All methods are evaluated under the same backbone, data partition, and training protocol for fair comparison. For \textsc{FedSGA}, the candidate split set $\mathcal{P}$ is selected from backbone block boundaries, and the reference split $p_{\mathrm{ref}}$ is set to the middle-depth candidate by default. We set the split-policy parameter to $\rho=0.1$, the shallow-sufficiency margin to $\beta=0.05$, and the prototype alignment weight to $\lambda=0.1$. The prototype temperature is set to $\tau_a=0.5$, and the prompt contribution temperature is set to $\tau_m=0.5$. We use $m=4$ client-specific prompt tokens by default and update the class prototypes with an EMA coefficient $\mu=0.9$. The calibration schedule $\mathcal{T}_{\mathrm{cal}}$ is set to perform calibration every 5 global rounds. The sufficiency estimator $\Gamma_\theta$ and the depth-aware projector are implemented as lightweight two-layer MLPs. 

\begin{table*}[t]
\centering
\caption{Accuracy comparison of \setlength{\fboxsep}{1pt}\colorbox{tablecolor}{\textsc{FedSGA}} and other benchmark methods on \underline{ResNet-18} backbone. The best accuracy is in \textbf{bold}.}
\renewcommand\arraystretch{0.8}
\label{overall_resnet}
\centering
\resizebox{1.0\textwidth}{!}{\begin{tabular}
{c|cc|cc|cc|cc} 
\toprule[1.2pt]

\multirow{2}{*}{\multirowcell{2}{\centering\textbf{Method}}} & \multicolumn{2}{c|}{\centering\textbf{CIFAR-10}} & \multicolumn{2}{c|}{\centering\textbf{CIFAR-100}} & \multicolumn{2}{c|}{\centering\textbf{Tiny-ImageNet}} & \multicolumn{2}{c}{\centering\textbf{DomainNet}}  \\ \cmidrule[0.5pt](l{1pt}r{0pt}){2-9}

& IID & Non-IID & IID & Non-IID & IID & Non-IID & IID & Non-IID \\ \cmidrule[0.8pt](l{1pt}r{0pt}){1-9}

\textsc{FedAvg} & 64.79 $\pm \scriptstyle{0.14}$ & 48.01 $\pm \scriptstyle{2.73}$ & 43.01 $\pm \scriptstyle{0.19}$ & 35.23 $\pm \scriptstyle{0.37}$ & 26.43 $\pm \scriptstyle{0.22}$ & 19.76 $\pm \scriptstyle{0.67}$ & 38.90 $\pm \scriptstyle{0.43}$ & 27.92 $\pm \scriptstyle{1.54}$ \\  

\textsc{FedProx} & 65.24 $\pm \scriptstyle{0.18}$ & 51.32 $\pm \scriptstyle{2.11}$ & 44.02 $\pm \scriptstyle{0.21}$ & 36.84 $\pm \scriptstyle{0.49}$ & 27.11 $\pm \scriptstyle{0.24}$ & 20.53 $\pm \scriptstyle{0.61}$ & 39.11 $\pm \scriptstyle{0.57}$ & 28.74 $\pm \scriptstyle{1.36}$  \\  

\textsc{FedBABU} & 67.88 $\pm \scriptstyle{0.24}$ & 57.84 $\pm \scriptstyle{1.76}$ & 47.23 $\pm \scriptstyle{0.28}$ & 40.18 $\pm \scriptstyle{0.57}$ & 29.84 $\pm \scriptstyle{0.31}$ & 23.96 $\pm \scriptstyle{0.74}$ & 41.76 $\pm \scriptstyle{0.34}$ & 32.45 $\pm \scriptstyle{1.12}$  \\  

\textsc{FedGH} & 68.57 $\pm \scriptstyle{0.21}$ & 59.21 $\pm \scriptstyle{1.44}$ & 48.32 $\pm \scriptstyle{0.25}$ & 41.36 $\pm \scriptstyle{0.46}$ & 30.62 $\pm \scriptstyle{0.27}$ & 24.88 $\pm \scriptstyle{0.69}$ & 42.38 $\pm \scriptstyle{0.39}$ & 33.27 $\pm \scriptstyle{0.98}$  \\  

\textsc{FedAS} & 70.45 $\pm \scriptstyle{0.17}$ & 61.47 $\pm \scriptstyle{1.28}$ & 50.67 $\pm \scriptstyle{0.45}$ & 43.92 $\pm \scriptstyle{0.41}$ & 32.72 $\pm \scriptstyle{0.25}$ & 27.15 $\pm \scriptstyle{0.58}$ & 44.59 $\pm \scriptstyle{0.17}$ & 35.64 $\pm \scriptstyle{0.86}$  \\  

\textsc{SplitFed} & 64.51 $\pm \scriptstyle{0.22}$ & 45.08 $\pm \scriptstyle{3.64}$ & 43.10 $\pm \scriptstyle{0.36}$ & 35.71 $\pm \scriptstyle{0.52}$ & 25.91 $\pm \scriptstyle{0.27}$ & 18.34 $\pm \scriptstyle{0.89}$ & 37.84 $\pm \scriptstyle{0.28}$ & 25.48 $\pm \scriptstyle{1.72}$   \\  

\textsc{MergeSFL} & 69.79 $\pm \scriptstyle{0.19}$ & 56.83 $\pm \scriptstyle{1.95}$ & 50.12 $\pm \scriptstyle{0.24}$ & 40.76 $\pm \scriptstyle{0.63}$ & 31.45 $\pm \scriptstyle{0.29}$ & 24.63 $\pm \scriptstyle{0.82}$ & 40.13 $\pm \scriptstyle{0.34}$ & 32.92 $\pm \scriptstyle{1.45}$  \\  

\textsc{FedMut} & 70.59 $\pm \scriptstyle{0.11}$ & 53.84 $\pm \scriptstyle{3.49}$ & 48.35 $\pm \scriptstyle{0.26}$ & 37.82 $\pm \scriptstyle{0.27}$ & 33.46 $\pm \scriptstyle{0.21}$ & 24.27 $\pm \scriptstyle{0.93}$ & 43.08 $\pm \scriptstyle{0.26}$ & 35.78 $\pm \scriptstyle{1.31}$ \\  

\textsc{MU-SplitFed} & 71.48 $\pm \scriptstyle{0.36}$ & 58.94 $\pm \scriptstyle{1.53}$ & 51.38 $\pm \scriptstyle{0.31}$ & 42.57 $\pm \scriptstyle{0.71}$ & 33.08 $\pm \scriptstyle{0.36}$ & 26.48 $\pm \scriptstyle{0.77}$ & 42.17 $\pm \scriptstyle{0.32}$ & 33.86 $\pm \scriptstyle{1.28}$  \\  

 \textsc{MultiSFL} & 73.19 $\pm \scriptstyle{0.13}$ & 66.72 $\pm \scriptstyle{0.48}$ & 56.06 $\pm \scriptstyle{0.16}$ & 47.17 $\pm \scriptstyle{0.19}$ & 38.72 $\pm \scriptstyle{0.18}$ & 33.41 $\pm \scriptstyle{0.36}$ & 48.35 $\pm \scriptstyle{0.16}$ & 40.26 $\pm \scriptstyle{0.54}$  \\  

\cellcolor{tablecolor}\textsc{FedSGA}(Ours) & \cellcolor{tablecolor}\textbf{79.22}$\pm \scriptstyle{0.18}$ & \cellcolor{tablecolor}\textbf{70.49}$\pm \scriptstyle{0.58}$ & \cellcolor{tablecolor}\textbf{62.44}$\pm \scriptstyle{0.21}$ & \cellcolor{tablecolor}\textbf{55.78}$\pm \scriptstyle{0.38}$ & \cellcolor{tablecolor}\textbf{43.89}$\pm \scriptstyle{1.39}$ & \cellcolor{tablecolor}\textbf{38.65}$\pm \scriptstyle{0.48}$ & \cellcolor{tablecolor}\textbf{54.13}$\pm \scriptstyle{0.26}$ & \cellcolor{tablecolor}\textbf{46.31}$\pm \scriptstyle{0.77}$ \\ 

\bottomrule[1.2pt]
\end{tabular}}
\label{accuracy_resnet}
\end{table*}

\begin{table*}[t]
\centering
\caption{Accuracy comparison of \setlength{\fboxsep}{1pt}\colorbox{tablecolor}{\textsc{FedSGA}} and other benchmark methods on \underline{ViT-B/16} backbone. The best accuracy is in \textbf{bold}.}
\renewcommand\arraystretch{0.8}
\label{overall_vit}
\centering
\resizebox{1.0\textwidth}{!}{\begin{tabular}
{c|cc|cc|cc|cc} 
\toprule[1.2pt]

\multirow{2}{*}{\multirowcell{2}{\centering\textbf{Method}}} & \multicolumn{2}{c|}{\centering\textbf{CIFAR-10}} & \multicolumn{2}{c|}{\centering\textbf{CIFAR-100}} & \multicolumn{2}{c|}{\centering\textbf{Tiny-ImageNet}} & \multicolumn{2}{c}{\centering\textbf{DomainNet}}  \\ \cmidrule[0.5pt](l{1pt}r{0pt}){2-9}

& IID & Non-IID & IID & Non-IID & IID & Non-IID & IID & Non-IID \\ \cmidrule[0.8pt](l{1pt}r{0pt}){1-9}

\textsc{FedAvg} & 72.84 $\pm \scriptstyle{0.12}$ & 56.31 $\pm \scriptstyle{2.14}$ & 49.72 $\pm \scriptstyle{0.18}$ & 40.96 $\pm \scriptstyle{0.41}$ & 32.88 $\pm \scriptstyle{0.26}$ & 24.37 $\pm \scriptstyle{0.58}$ & 44.62 $\pm \scriptstyle{0.37}$ & 32.85 $\pm \scriptstyle{1.42}$ \\

\textsc{FedProx} & 73.56 $\pm \scriptstyle{0.15}$ & 59.48 $\pm \scriptstyle{1.83}$ & 50.83 $\pm \scriptstyle{0.21}$ & 42.37 $\pm \scriptstyle{0.46}$ & 33.71 $\pm \scriptstyle{0.22}$ & 25.68 $\pm \scriptstyle{0.54}$ & 45.21 $\pm \scriptstyle{0.42}$ & 34.16 $\pm \scriptstyle{1.27}$ \\

\textsc{FedBABU} & 76.42 $\pm \scriptstyle{0.19}$ & 64.37 $\pm \scriptstyle{1.44}$ & 54.86 $\pm \scriptstyle{0.24}$ & 46.83 $\pm \scriptstyle{0.51}$ & 37.12 $\pm \scriptstyle{0.28}$ & 30.41 $\pm \scriptstyle{0.63}$ & 48.93 $\pm \scriptstyle{0.31}$ & 38.85 $\pm \scriptstyle{1.04}$ \\

\textsc{FedGH} & 77.28 $\pm \scriptstyle{0.17}$ & 65.91 $\pm \scriptstyle{1.26}$ & 55.73 $\pm \scriptstyle{0.23}$ & 48.15 $\pm \scriptstyle{0.43}$ & 38.26 $\pm \scriptstyle{0.24}$ & 31.74 $\pm \scriptstyle{0.57}$ & 49.84 $\pm \scriptstyle{0.28}$ & 40.12 $\pm \scriptstyle{0.92}$ \\

\textsc{FedAS} & 79.35 $\pm \scriptstyle{0.14}$ & 68.42 $\pm \scriptstyle{1.03}$ & 58.96 $\pm \scriptstyle{0.31}$ & 50.93 $\pm \scriptstyle{0.36}$ & 41.08 $\pm \scriptstyle{0.21}$ & 35.46 $\pm \scriptstyle{0.44}$ & 52.77 $\pm \scriptstyle{0.16}$ & 43.56 $\pm \scriptstyle{0.73}$ \\

\textsc{SplitFed} & 72.11 $\pm \scriptstyle{0.18}$ & 53.26 $\pm \scriptstyle{3.08}$ & 49.15 $\pm \scriptstyle{0.29}$ & 40.42 $\pm \scriptstyle{0.48}$ & 31.94 $\pm \scriptstyle{0.25}$ & 23.18 $\pm \scriptstyle{0.77}$ & 43.26 $\pm \scriptstyle{0.24}$ & 30.74 $\pm \scriptstyle{1.58}$ \\

\textsc{MergeSFL} & 78.36 $\pm \scriptstyle{0.16}$ & 63.88 $\pm \scriptstyle{1.62}$ & 57.84 $\pm \scriptstyle{0.22}$ & 47.65 $\pm \scriptstyle{0.57}$ & 39.67 $\pm \scriptstyle{0.27}$ & 31.28 $\pm \scriptstyle{0.71}$ & 47.58 $\pm \scriptstyle{0.29}$ & 39.47 $\pm \scriptstyle{1.18}$ \\

\textsc{FedMut} & 79.61 $\pm \scriptstyle{0.10}$ & 60.72 $\pm \scriptstyle{2.94}$ & 56.38 $\pm \scriptstyle{0.25}$ & 44.26 $\pm \scriptstyle{0.24}$ & 40.95 $\pm \scriptstyle{0.18}$ & 30.86 $\pm \scriptstyle{0.85}$ & 51.26 $\pm \scriptstyle{0.21}$ & 42.38 $\pm \scriptstyle{1.12}$ \\

\textsc{MU-SplitFed} & 80.42 $\pm \scriptstyle{0.27}$ & 65.48 $\pm \scriptstyle{1.31}$ & 59.02 $\pm \scriptstyle{0.28}$ & 49.13 $\pm \scriptstyle{0.63}$ & 41.63 $\pm \scriptstyle{0.31}$ & 34.12 $\pm \scriptstyle{0.69}$ & 50.18 $\pm \scriptstyle{0.27}$ & 42.15 $\pm \scriptstyle{1.06}$ \\

\textsc{MultiSFL} & 82.73 $\pm \scriptstyle{0.09}$ & 73.18 $\pm \scriptstyle{0.39}$ & 64.47 $\pm \scriptstyle{0.14}$ & 55.82 $\pm \scriptstyle{0.17}$ & 47.91 $\pm \scriptstyle{0.16}$ & 41.26 $\pm \scriptstyle{0.31}$ & 56.94 $\pm \scriptstyle{0.13}$ & 48.72 $\pm \scriptstyle{0.47}$ \\

\cellcolor{tablecolor}\textsc{FedSGA}(Ours) & \cellcolor{tablecolor}\textbf{86.58}$\pm \scriptstyle{0.16}$ & \cellcolor{tablecolor}\textbf{78.93}$\pm \scriptstyle{0.67}$ & \cellcolor{tablecolor}\textbf{68.94}$\pm \scriptstyle{0.23}$ & \cellcolor{tablecolor}\textbf{59.78}$\pm \scriptstyle{0.31}$ & \cellcolor{tablecolor}\textbf{53.36}$\pm \scriptstyle{0.24}$ & \cellcolor{tablecolor}\textbf{48.72}$\pm \scriptstyle{0.53}$ & \cellcolor{tablecolor}\textbf{61.09}$\pm \scriptstyle{0.26}$ & \cellcolor{tablecolor}\textbf{54.65}$\pm \scriptstyle{0.49}$ \\ 

\bottomrule[1.2pt]
\end{tabular}}
\label{accuracy_vit}
\end{table*}

\subsection{Main Results and Analysis (RQ1)}
Tables~\ref{overall_resnet} and~\ref{overall_vit} report the overall accuracy comparison under ResNet-18 and ViT-B/16 backbones. \textsc{FedSGA} consistently achieves the best performance across all datasets, data partitions, and backbone architectures, demonstrating the effectiveness of sufficiency-guided adaptive split selection under heterogeneous client conditions. Compared with conventional FL and SFL baselines, \textsc{FedSGA} shows more pronounced advantages under Non-IID settings, where clients exhibit stronger distributional and representation-level discrepancies. This suggests that simply aggregating client updates or adopting a fixed split interface is insufficient when client-side representations evolve at different rates. The improvement is particularly evident on more challenging datasets such as Tiny-ImageNet and DomainNet, where heterogeneous semantics and complex visual domains make split-interface reliability harder to assess. In these cases, baselines relying on static partitions, resource-oriented split decisions, or post hoc representation alignment tend to suffer greater performance degradation. In contrast, \textsc{FedSGA} explicitly estimates whether the shallow interface is already sufficient for each client and further harmonizes heterogeneous-depth activations before server-side prediction, leading to more stable collaborative training. The consistent gains on both ResNet-18 and ViT-B/16 further indicate that the proposed framework is not tied to a specific backbone architecture and remains effective for both convolutional and transformer-based models, providing a general mechanism for improving adaptive split learning under client heterogeneity.

\begin{table*}[t]
\centering
\renewcommand\arraystretch{0.8}
\caption{Efficacy of each module on various datasets.}
\label{ablation}
\vspace{-10pt}
\begin{subtable}{\linewidth}
\centering
\caption{ResNet-18}
\resizebox{1.0\textwidth}{!}{
\begin{tabular}
{ccc|c|c|c|c|c|c} 
\toprule[1.2pt]

\multicolumn{3}{c|}{\centering\textbf{Module}} & \multicolumn{3}{c|}{\centering\textbf{IID}} & \multicolumn{3}{c}{\centering\textbf{Non-IID}}    \\ \cmidrule[0.5pt](l{1pt}r{0pt}){1-9}

\textbf{Adapt.} & \textbf{Suff. Est.} & \textbf{Harmon.}  & \textbf{CIFAR-100} & \textbf{Tiny-ImageNet} & \textbf{DomainNet} & \textbf{CIFAR-100} & \textbf{Tiny-ImageNet} & \textbf{DomainNet}    \\ \cmidrule[0.5pt](l{1pt}r{0pt}){1-9}

&  &  & 43.10 $\pm \scriptstyle{0.36}$ & 25.91 $\pm \scriptstyle{0.27}$ & 37.84 $\pm \scriptstyle{0.28}$ & 35.71 $\pm \scriptstyle{0.52}$ & 18.34 $\pm \scriptstyle{0.89}$ & 25.48 $\pm \scriptstyle{1.72}$  \\ 

& \textcolor{olive}{\ding{52}} & \textcolor{olive}{\ding{52}}  & 59.84 $\pm \scriptstyle{0.24}$ & 40.61 $\pm \scriptstyle{1.21}$ & 50.47 $\pm \scriptstyle{0.35}$ & 51.36 $\pm \scriptstyle{0.46}$ & 34.72 $\pm \scriptstyle{0.61}$ & 41.28 $\pm \scriptstyle{0.94}$  \\

\textcolor{olive}{\ding{52}} &  & \textcolor{olive}{\ding{52}} & 56.73 $\pm \scriptstyle{0.28}$ & 37.82 $\pm \scriptstyle{1.16}$ & 47.66 $\pm \scriptstyle{0.31}$ & 47.38 $\pm \scriptstyle{0.55}$ & 31.24 $\pm \scriptstyle{0.73}$ & 38.64 $\pm \scriptstyle{1.02}$  \\

\textcolor{olive}{\ding{52}} & \textcolor{olive}{\ding{52}} &  & 58.91 $\pm \scriptstyle{0.26}$ & 39.47 $\pm \scriptstyle{1.28}$ & 48.92 $\pm \scriptstyle{0.38}$ & 50.14 $\pm \scriptstyle{0.49}$ & 33.56 $\pm \scriptstyle{0.64}$ & 40.37 $\pm \scriptstyle{0.88}$  \\ \midrule[0.8pt]

\textcolor{olive}{\ding{52}} & \textcolor{olive}{\ding{52}} & \textcolor{olive}{\ding{52}} & \textbf{62.44}$\pm \scriptstyle{0.21}$ & \textbf{43.89}$\pm \scriptstyle{1.39}$ & \textbf{54.13}$\pm \scriptstyle{0.26}$ & \textbf{55.78}$\pm \scriptstyle{0.38}$ & \textbf{38.65}$\pm \scriptstyle{0.48}$ & \textbf{46.31}$\pm \scriptstyle{0.77}$ \\ 

\bottomrule[1.2pt]
\end{tabular}}
\label{ablation_resnet}
\end{subtable}

\vspace{3pt}  

\begin{subtable}{\linewidth}
\centering
\caption{ViT-B/16}
\resizebox{1.0\textwidth}{!}{\begin{tabular}
{ccc|c|c|c|c|c|c} 
\toprule[1.2pt]

\multicolumn{3}{c|}{\centering\textbf{Module}} & \multicolumn{3}{c|}{\centering\textbf{IID}} & \multicolumn{3}{c}{\centering\textbf{Non-IID}}    \\ \cmidrule[0.5pt](l{1pt}r{0pt}){1-9}

\textbf{Adapt.} & \textbf{Suff. Est.} & \textbf{Harmon.}  & \textbf{CIFAR-100} & \textbf{Tiny-ImageNet} & \textbf{DomainNet} & \textbf{CIFAR-100} & \textbf{Tiny-ImageNet} & \textbf{DomainNet}    \\ \cmidrule[0.5pt](l{1pt}r{0pt}){1-9}

&  &  & 49.15 $\pm \scriptstyle{0.29}$ & 31.94 $\pm \scriptstyle{0.25}$ & 43.26 $\pm \scriptstyle{0.24}$ & 40.42 $\pm \scriptstyle{0.48}$ & 23.18 $\pm \scriptstyle{0.77}$ & 30.74 $\pm \scriptstyle{1.58}$  \\ 

& \textcolor{olive}{\ding{52}} & \textcolor{olive}{\ding{52}}  & 66.24 $\pm \scriptstyle{0.26}$ & 50.18 $\pm \scriptstyle{0.31}$ & 57.42 $\pm \scriptstyle{0.34}$ & 56.21 $\pm \scriptstyle{0.39}$ & 44.83 $\pm \scriptstyle{0.58}$ & 50.16 $\pm \scriptstyle{0.61}$  \\

\textcolor{olive}{\ding{52}} &  & \textcolor{olive}{\ding{52}} & 63.87 $\pm \scriptstyle{0.29}$ & 47.36 $\pm \scriptstyle{0.37}$ & 54.69 $\pm \scriptstyle{0.31}$ & 52.48 $\pm \scriptstyle{0.43}$ & 41.62 $\pm \scriptstyle{0.66}$ & 46.72 $\pm \scriptstyle{0.74}$  \\

\textcolor{olive}{\ding{52}} & \textcolor{olive}{\ding{52}} &  & 65.31 $\pm \scriptstyle{0.25}$ & 49.27 $\pm \scriptstyle{0.34}$ & 55.84 $\pm \scriptstyle{0.36}$ & 54.83 $\pm \scriptstyle{0.41}$ & 43.76 $\pm \scriptstyle{0.61}$ & 48.38 $\pm \scriptstyle{0.68}$  \\ \midrule[0.8pt]

\textcolor{olive}{\ding{52}} & \textcolor{olive}{\ding{52}} & \textcolor{olive}{\ding{52}} & \textbf{68.94}$\pm \scriptstyle{0.23}$ & \textbf{53.36}$\pm \scriptstyle{0.24}$ & \textbf{61.09}$\pm \scriptstyle{0.26}$ & \textbf{59.78}$\pm \scriptstyle{0.31}$ & \textbf{48.72}$\pm \scriptstyle{0.53}$ & \textbf{54.65}$\pm \scriptstyle{0.49}$  \\ 

\bottomrule[1.2pt]
\end{tabular}}
\label{ablation_vit}
\end{subtable}
\end{table*}

\begin{figure*}[t]
\centering
\subfloat{
    \includegraphics[width=0.24\textwidth, trim= 5 5 5 5,clip]{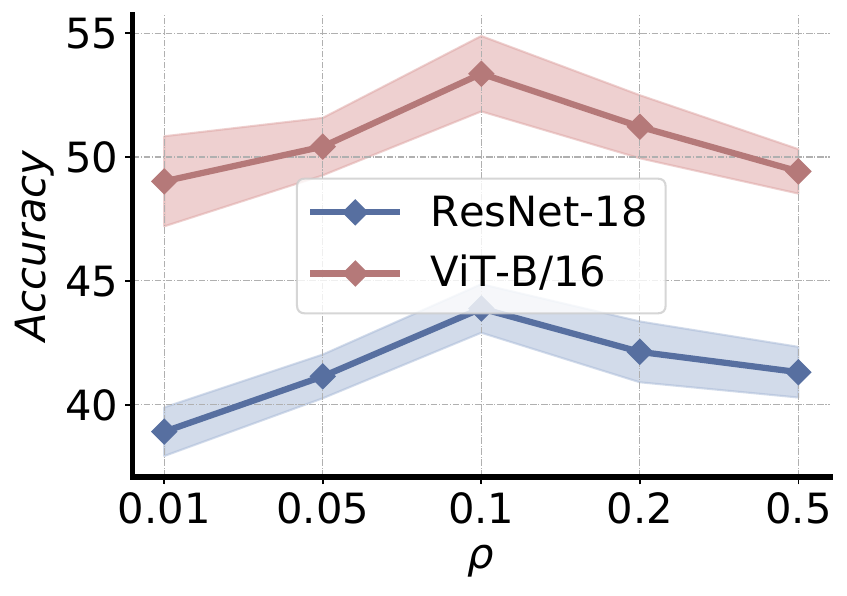}}
\subfloat{
    \includegraphics[width=0.24\textwidth, trim=5 5 5 5,clip]{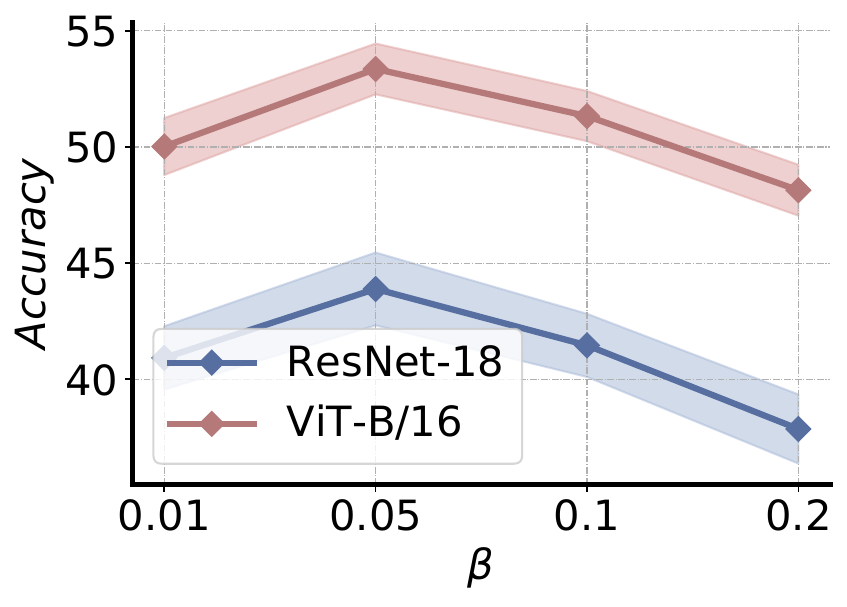}}
\subfloat{
    \includegraphics[width=0.24\textwidth, trim=5 5 5 5,clip]{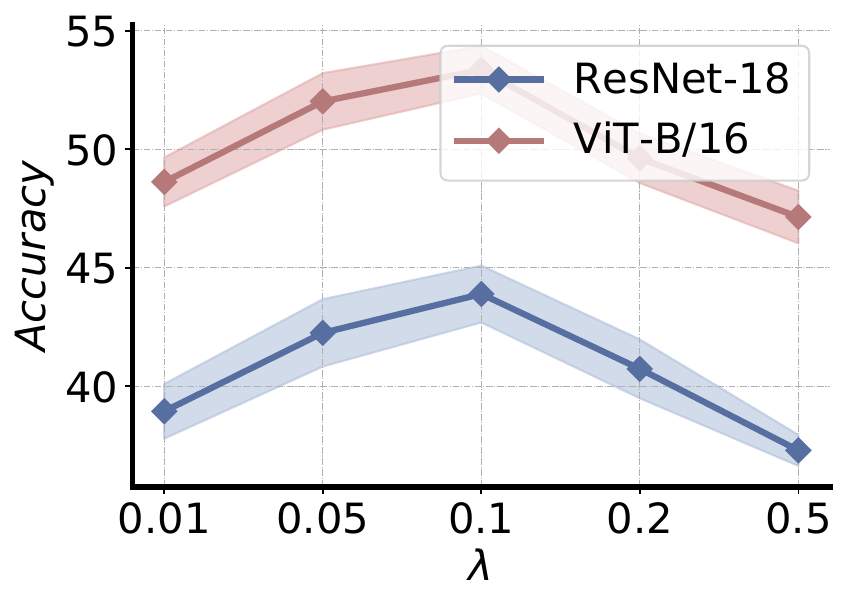}}
\subfloat{
    \includegraphics[width=0.24\textwidth, trim=5 5 5 5,clip]{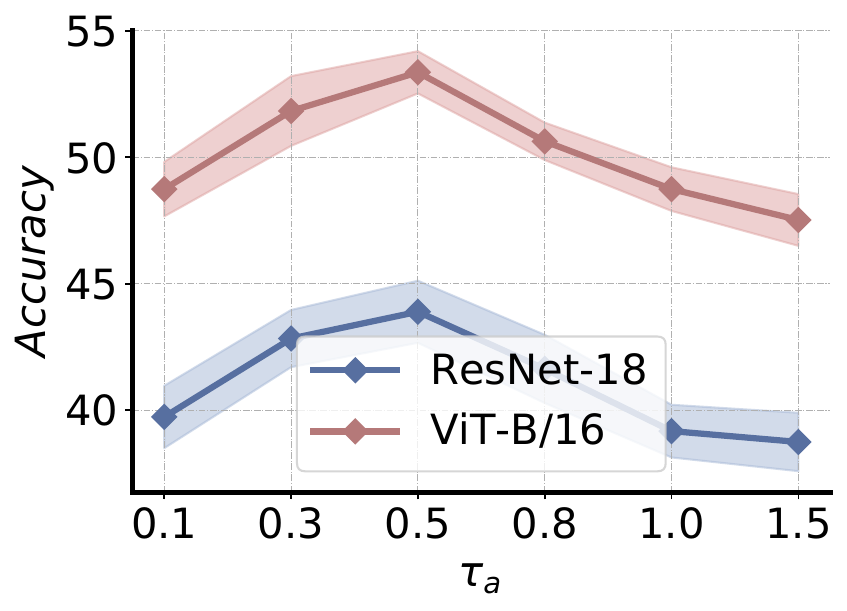}}
\vspace{-8pt}
\caption{Impact of hyperparameters on model performance on the CIFAR-100 dataset.} 
\label{hyper_cifar100}
\end{figure*}

\begin{figure*}[t]
\centering
\subfloat{
    \includegraphics[width=0.24\textwidth, trim= 5 5 5 5,clip]{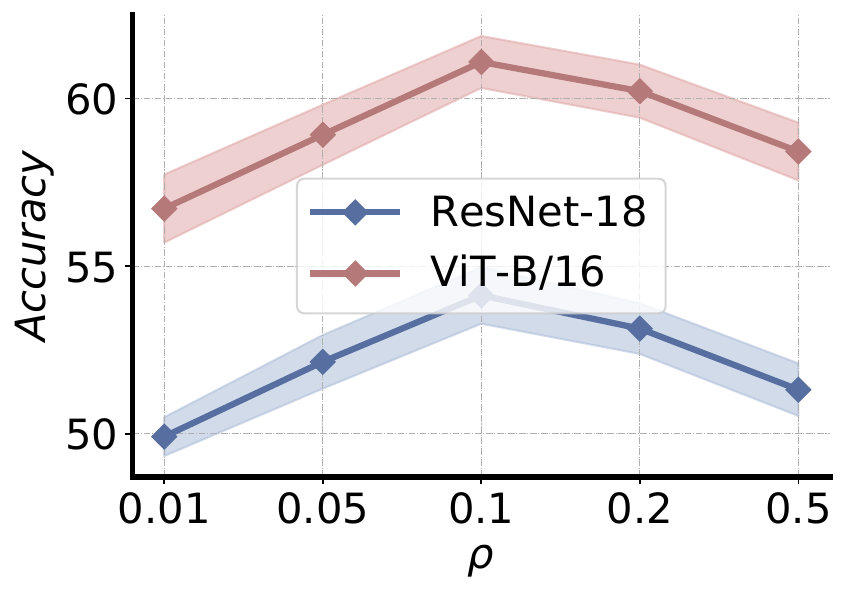}}
\subfloat{
    \includegraphics[width=0.24\textwidth, trim=5 5 5 5,clip]{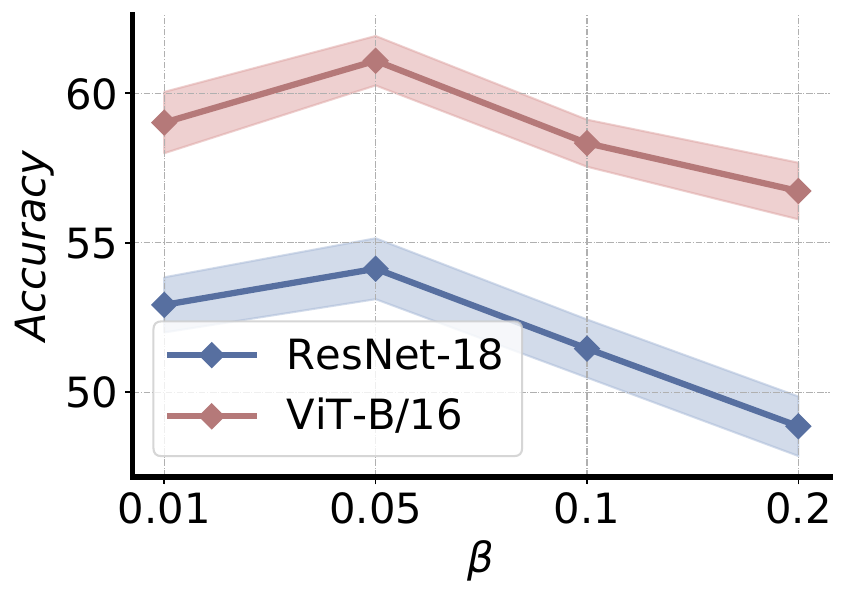}}
\subfloat{
    \includegraphics[width=0.24\textwidth, trim=5 5 5 5,clip]{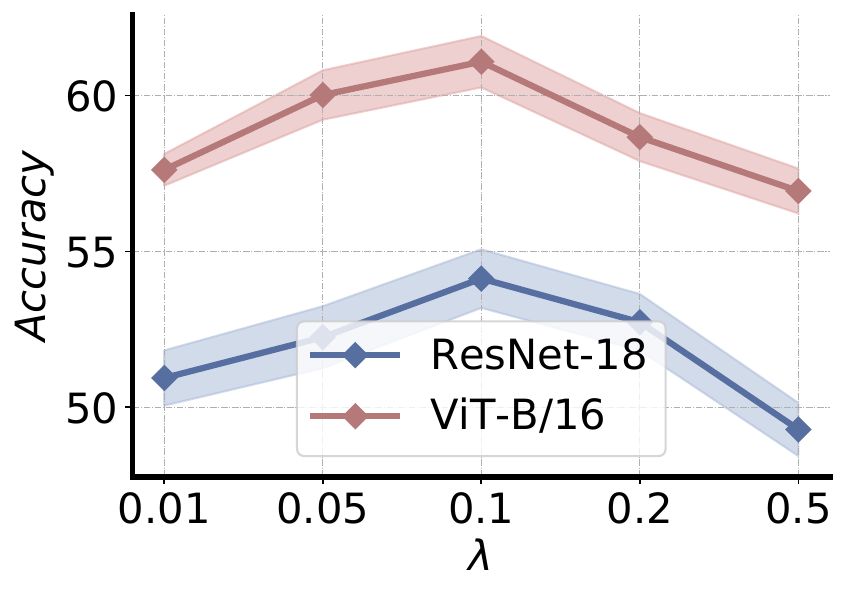}}
\subfloat{
    \includegraphics[width=0.24\textwidth, trim=5 5 5 5,clip]{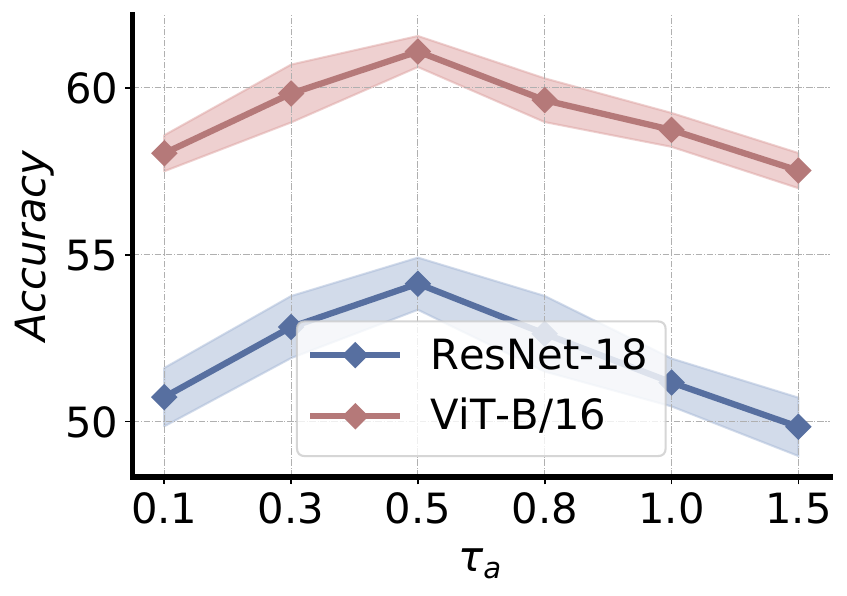}}
\vspace{-8pt}
\caption{Impact of hyperparameters on model performance on the DomainNet dataset.} 
\label{hyper_domainnet}
\end{figure*}

\subsection{Ablation Study (RQ2)}
We provide the ablation study of the three main components in Table~\ref{ablation}, including the client-specific adaptation channel (\textbf{Adapt.}), the shallow sufficiency estimator (\textbf{Suff. Est.}), and the split-compatible interface harmonization module (\textbf{Harmon.}). The results consistently show that all three components contribute positively across datasets, data partitions, and backbone architectures. The full model achieves the best performance in all cases, indicating that reliable adaptive split learning requires both client-local adaptation awareness and split-compatible server-side interface modeling.

Among the three components, the shallow sufficiency estimator plays a central role in split selection. Removing \textbf{Suff. Est.} leads to consistent performance drops under both ResNet-18 and ViT-B/16, showing that adaptive splitting without explicitly estimating shallow-interface sufficiency can result in unreliable split decisions. The client-specific adaptation channel also provides stable gains, especially under Non-IID and DomainNet settings, where local adaptation dynamics are more heterogeneous. This supports our motivation that prompt-state variation captures residual client-local adaptation that cannot be fully reflected by instantaneous representation quality alone. The harmonization module further improves performance by making activations from different split depths more comparable before server-side prediction. Without \textbf{Harmon.}, accuracy consistently decreases, confirming that heterogeneous split depths introduce interface mismatch that cannot be handled by adaptive split selection alone.

\begin{table*}[t]
\centering
\renewcommand\arraystretch{0.8}
\caption{Computational cost analysis under different backbone architectures.}
\label{communication}
\vspace{-10pt}
\begin{subtable}{\linewidth}
\centering
\caption{ResNet-18}
\resizebox{1.0\textwidth}{!}{
\begin{tabular}
{c|cccc|cccc|cccc} 
\toprule[1.2pt]

\multirow{1.5}{*}{\multirowcell{2}{\textbf{Method}}} & \multicolumn{4}{c|}{\centering\textbf{CIFAR-100}} & \multicolumn{4}{c|}{\centering\textbf{Tiny-ImageNet}} & \multicolumn{4}{c}{\centering\textbf{DomainNet}}  \\ \cmidrule[0.5pt](l{1pt}r{0pt}){2-13}

& Params & FLOPs & Time(s) & Acc. & Params & FLOPs & Time(s) & Acc. & Params & FLOPs & Time(s) & Acc. \\ \cmidrule[0.8pt](l{1pt}r{0pt}){1-13}



\textsc{MultiSFL} & 5.58M & 0.62G & 41.27 & 56.06 & 5.64M & 1.84G & 78.36 & 38.72 & 5.71M & 3.92G & 116.48 & 48.35  \\

\cellcolor{tablecolor}\textsc{FedSGA} & \cellcolor{tablecolor}4.31M & \cellcolor{tablecolor}0.49G & \cellcolor{tablecolor}36.14 & \cellcolor{tablecolor}62.44 & \cellcolor{tablecolor}4.72M & \cellcolor{tablecolor}1.55G & \cellcolor{tablecolor}68.92 & \cellcolor{tablecolor}43.89 & \cellcolor{tablecolor}4.96M & \cellcolor{tablecolor}3.28G & \cellcolor{tablecolor}101.37 & \cellcolor{tablecolor}54.13  \\

\bottomrule[1.2pt]
\end{tabular}}
\label{overhead_resnet}
\end{subtable}

\vspace{3pt}  

\begin{subtable}{\linewidth}
\centering
\caption{ViT-B/16}
\resizebox{1.0\textwidth}{!}{
\begin{tabular}
{c|cccc|cccc|cccc} 
\toprule[1.2pt]

\multirow{1.5}{*}{\multirowcell{2}{\textbf{Method}}} & \multicolumn{4}{c|}{\centering\textbf{CIFAR-100}} & \multicolumn{4}{c|}{\centering\textbf{Tiny-ImageNet}} & \multicolumn{4}{c}{\centering\textbf{DomainNet}}  \\ \cmidrule[0.5pt](l{1pt}r{0pt}){2-13}

& Params & FLOPs & Time(s) & Acc. & Params & FLOPs & Time(s) & Acc. & Params & FLOPs & Time(s) & Acc. \\ \cmidrule[0.8pt](l{1pt}r{0pt}){1-13}



\textsc{MultiSFL} & 42.63M & 8.74G & 96.41 & 64.47 & 42.71M & 13.86G & 142.73 & 47.91 & 42.83M & 18.92G & 205.36 & 56.94   \\

\cellcolor{tablecolor}\textsc{FedSGA} & \cellcolor{tablecolor}34.18M & \cellcolor{tablecolor}6.91G & \cellcolor{tablecolor}84.25 & \cellcolor{tablecolor}68.94 & \cellcolor{tablecolor}36.74M & \cellcolor{tablecolor}11.27G & \cellcolor{tablecolor}126.58 & \cellcolor{tablecolor}53.36 & \cellcolor{tablecolor}38.95M & \cellcolor{tablecolor}15.64G & \cellcolor{tablecolor}181.42 & \cellcolor{tablecolor}61.09     \\

\bottomrule[1.2pt]
\end{tabular}}
\label{overhead_vit}
\end{subtable}
\end{table*}

\subsection{Hyperparameter Sensitivity Analysis (RQ3)}
Figures~\ref{hyper_cifar100} and~\ref{hyper_domainnet} illustrate the impact of key hyperparameters on \textsc{FedSGA} on the CIFAR-100 and DomainNet datasets, respectively, using both ResNet-18 and ViT-B/16 backbones. For the split policy parameter $\rho$, the best performance is generally obtained around $\rho=0.1$. A smaller $\rho$ favors earlier offloading, which may select shallow interfaces too aggressively before they become sufficiently reliable. In contrast, a larger $\rho$ shifts the policy toward deeper client-side computation, reducing the benefit of adaptive shallow splitting and potentially weakening the efficiency-performance trade-off. For the sufficiency margin $\beta$, moderate values achieve better results, with $\beta=0.05$ performing best in most cases. This suggests that an overly strict sufficiency criterion may reject useful shallow interfaces, whereas an overly relaxed criterion may accept premature shallow splits. The alignment weight $\lambda$ also shows a clear optimal range. Performance improves as $\lambda$ increases from a small value, but degrades when $\lambda$ becomes too large. This indicates that prototype-based alignment is beneficial for harmonizing heterogeneous split interfaces, but excessive alignment may over-constrain the task representation. Similarly, the temperature $\tau_a$ achieves the best performance around $\tau_a=0.5$. Too small a temperature can make prototype alignment overly sharp, while too large a temperature weakens class-level discrimination in the shared semantic space. Overall, these results show that \textsc{FedSGA} benefits from a balanced configuration for sufficiency estimation and interface harmonization, while maintaining stable performance across datasets and backbone architectures.

\begin{figure}[t]
\centering
\subfloat[CIFAR-100]{
    \includegraphics[width=0.23\textwidth, trim=30 30 70 40,clip]{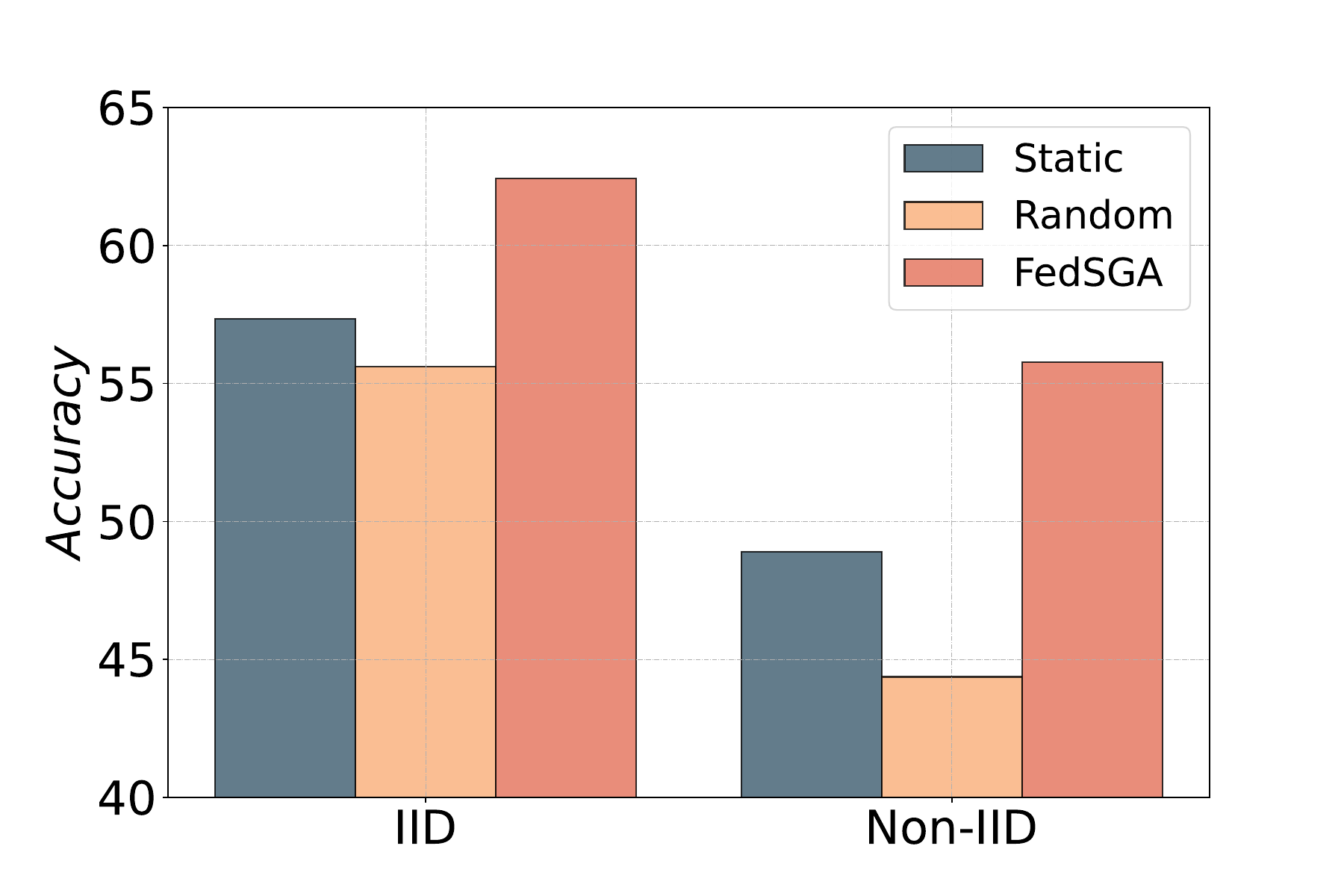}}
\subfloat[DomainNet]{
    \includegraphics[width=0.23\textwidth, trim=30 30 70 40,clip]{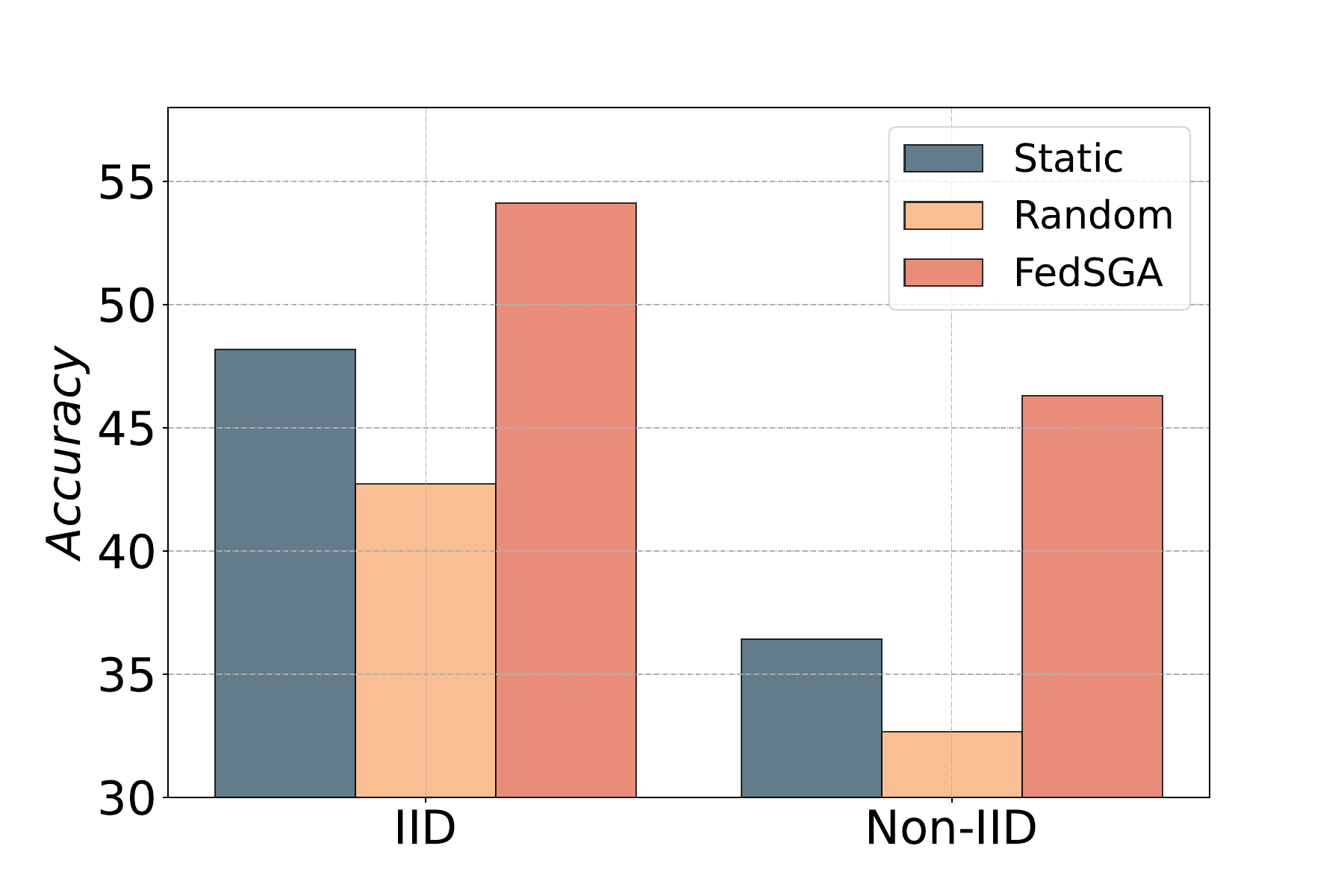}} 
\caption{Comparison of different split strategies on the CIFAR-100 and DomainNet datasets.}
\label{split}
\end{figure}

\subsection{Computational Efficiency Analysis (RQ4)}
In Table~\ref{communication}, we present the efficiency-performance trade-off of \textsc{FedSGA}. Compared with \textsc{MultiSFL}, \textsc{FedSGA} consistently achieves higher accuracy while reducing client-side computational cost across all datasets. Specifically, \textsc{FedSGA} requires fewer activated parameters, lower FLOPs, and shorter training time, indicating that the proposed sufficiency-guided adaptive split strategy can avoid unnecessary deep local computation without sacrificing representation quality. The efficiency gains become more evident on larger-scale and more heterogeneous datasets such as DomainNet, where \textsc{FedSGA} consistently reduces computational overhead while maintaining clear accuracy improvements. This behavior is consistent with the design motivation of the proposed framework. Instead of using a fixed split strategy for all clients, \textsc{FedSGA} dynamically adjusts the split depth according to the estimated shallow-interface sufficiency and client-local adaptation state. As a result, clients whose shallow representations are already reliable can offload earlier, while clients undergoing stronger local adaptation can retain deeper local computation when necessary. The results therefore demonstrate that the proposed adaptive split mechanism achieves a more favorable balance between computational efficiency and collaborative learning performance under heterogeneous federated environments.

\subsection{Split Strategy Comparison (RQ5)}
Figure~\ref{split} compares \textsc{FedSGA} with two alternative split strategies, including a fixed static split and random split selection, on CIFAR-100 and DomainNet. \textsc{FedSGA} consistently achieves the highest accuracy under both IID and Non-IID settings, demonstrating the effectiveness of sufficiency-guided split selection. Compared with the static strategy, \textsc{FedSGA} improves performance by selecting client-specific split depths rather than enforcing a uniform partition for all clients. Compared with random splitting, the large performance gap shows that adaptive split selection must be guided by reliable interface sufficiency signals instead of arbitrary depth choices. The improvement is particularly clear under Non-IID settings. On CIFAR-100 and DomainNet, \textsc{FedSGA} substantially outperforms both static and random strategies, indicating that client heterogeneity makes fixed or unguided split decisions unreliable. Static splitting cannot accommodate clients whose representations mature at different depths, while random splitting may frequently produce premature or incompatible interfaces. In contrast, \textsc{FedSGA} estimates shallow sufficiency from semantic alignment, temporal stability, and prompt-state variation, allowing each client to offload only when its selected interface is likely to be reliable. These results validate that the gains of \textsc{FedSGA} come from principled client-specific split decisions rather than simply varying the split point.

\section{Conclusion}
In this paper, we proposed \textsc{FedSGA}, a sufficiency-guided adaptive split federated learning framework for heterogeneous clients. Instead of relying on a uniform static split point or repeatedly probing multiple candidate depths, \textsc{FedSGA} formulates split selection as a client-specific shallow sufficiency estimation problem. To capture whether client-local adaptation remains active, we introduced a private prompt-based adaptation channel and used prompt-state variation as a lightweight signal complementary to semantic alignment and temporal interface stability. Based on these signals, \textsc{FedSGA} estimates whether the shallowest interface is already sufficient and selects the split depth through a budget-aware monotone policy. To support clients offloading from different depths, we further developed a split-compatible harmonization module that maps heterogeneous interface activations into a shared semantic space with prototype-based regularization. Extensive experiments on multiple heterogeneous benchmarks and backbone architectures demonstrated that \textsc{FedSGA} consistently improves model performance over state-of-the-art methods while reducing client-side computation, validating the effectiveness of sufficiency-guided adaptive split learning under client heterogeneity.

\section*{Acknowledge}
This work was supported by the UGC General Research Fund no. 17209822 and the Innovation and Technology Commission Fund no. ITS/383/23FP from Hong Kong.

\newpage

\section*{GenAI Usage Disclosure}
The above work is entirely our original work, without any content generated by generative AI tools. 


\bibliographystyle{ACM-Reference-Format}
\bibliography{reference}










\end{document}